\documentclass[preprint, 11pt]{aastex}
\shorttitle{NGC1614 in High Resolution}
\shortauthors{Sliwa et al.}

\usepackage{float}

\newcommand{\tco}{$^{13}$CO}
\newcommand{\co}{$^{12}$CO}
\newcommand{\tcoone}{$^{13}$CO $J$=1-0}
\newcommand{\coone}{$^{12}$CO $J$=1-0}
\newcommand{\cotwo}{$^{12}$CO $J$=2-1}
\newcommand{\tcotwo}{$^{13}$CO $J$=2-1}
\newcommand{\cothree}{$^{12}$CO $J$=3-2}
\newcommand{\cosix}{$^{12}$CO $J$=6-5}
\newcommand{\lfir}{$L_{\rm{FIR}}$}
\newcommand{\lsol}{$L_{\odot}$}
\newcommand{\msol}{$M_{\odot}$}
\newcommand{\mdyn}{$M_{dyn}$}

\newcommand{\kms}{km s$^{-1}$}
\newcommand{\hcofour}{HCO$^+$ $J$=4$-$3}
\newcommand{\tkin}{$T_{\rm{kin}}$} 
\newcommand{\nhtwo}{$n_{\rm{H_{2}}}$}
\newcommand{\alphaco}{$M_{\odot}$ (K km s$^{-1}$ pc$^{2}$)$^{-1}$}
\newcommand{\nco}{$N_{\rm{^{12}CO}}$}
\newcommand{\ff}{$\Phi_{\rm{A}}$}

\newcommand{\xco}{[$^{12}$CO]/[$^{13}$CO]}

\newcommand{\xh}{$x_{\rm{CO}}$}

\begin{document}

\title{Around the Ring We Go: The Cold, Dense Ring of Molecular Gas in NGC 1614}
\author{Kazimierz Sliwa\altaffilmark{1},  
	Christine D. Wilson\altaffilmark{1},
            Daisuke Iono\altaffilmark{2},
	 Alison Peck\altaffilmark{3},
          Satoki Matsushita\altaffilmark{4}
}
 
\altaffiltext{1}{Department of Physics and Astronomy, McMaster University, Hamilton, ON L8S 4M1 Canada; sliwak@mcmaster.ca, wilson@physics.mcmaster.ca}
\altaffiltext{2}{National Astronomical Observatory of Japan, 2-21-1, Osawa, Mitaka, Tokyo 181-8588 d.iono@nao.ac.jp}
\altaffiltext{3}{National Radio Astronomy Observatory, Charlottesville, VA 22903, USA; apeck@alma.cl}
\altaffiltext{4}{Academia Sinica Institute of Astronomy and Astrophysics, P.O. Box 23-141, Taipei 10617, Taiwan; satoki@asiaa.sinica.edu.tw}

\begin{abstract}
We present high-resolution archival Atacama Large Millimeter/submillimeter Array (ALMA) \cothree\ and $J$=6-5 and \hcofour\ observations and new CARMA \co\ and \tcoone\ observations of the luminous infrared galaxy NGC 1614. The high-resolution maps show the previously identified ring-like structure while the \cothree\ map shows extended emission that traces the extended dusty features. We combined these new observations with previously published Submillimeter Array \co\ and \tcotwo\ observations to constrain the physical conditions of the molecular gas at a resolution of 230 pc using a radiative transfer code and a Bayesian likelihood analysis. At several positions around the central ring-like structure, the molecular gas is cold (20-40 K) and dense ($>$ 10$^{3.0}$ cm$^{-3}$) . The only region that shows evidence of a second molecular gas component is the ``hole" in the ring. The \co-to-\tco\ abundance ratio is found to be greater than 130, more than twice the local interstellar medium value. We also measure the CO-to-H$_{2}$ conversion factor, $\alpha_{CO}$, to range from 0.9 to 1.5 \alphaco. 
\end{abstract}

\keywords{galaxies:  individual(NGC 1614, Arp 186, Mrk 0617, IRAS 04315-0840, II Zw 015) ---  galaxies: interactions --- galaxies: starburst --- galaxies: abundances --- submillimeter: galaxies --- radiative transfer}

\section{Introduction} 

High-resolution observations of molecular gas in nearby luminous infrared galaxies (LIRGs) reveal concentrations of gas near the nuclei of the merging galaxies (e.g. Downes $\&$ Solomon 1998\nocite{Downes1998}, Wilson et al. 2008\nocite{Wilson2008}).  Tidal interactions of gas-rich galaxies are largely responsible for funnelling the molecular gas to the central kiloparsec, compressing the gas which, in turn, triggers a starburst (e.g. Mihos $\&$ Hernquist 1996\nocite{Mihos1996}). Even minor mergers between a gas-rich galaxy and a satellite companion can be efficient at driving molecular gas towards the central regions (e.g. Mihos $\&$ Hernquist 1994\nocite{Mihos1994}).

NGC 1614 ($D_{L}$ = 67.9 Mpc; 1\arcsec\ = 330 pc) has a far-infrared luminosity (\lfir) of 2.7 $\times$ 10$^{11}$ \lsol\ (Sanders et al. 2003)\nocite{Sanders2003}. Optically, this galaxy has a bright center and two fairly symmetric inner spiral arms \citep{Neff1990}. A circumnuclear ring is seen in several tracers, such as Pa$\alpha$ \citep{AlonsoHerrero2001}, radio continuum \citep{Olsson2010}, PAHs \citep{Vaisanen2012} and \cotwo\ \citep{Konig2013}. The ring hosts a very young starburst (5-10 Myr), while the nucleus hosts an older starburst ($>$10 Myr; Olsson et al. 2010).  Using the Atacama Large Millimeter/submillimeter Array (ALMA), Imanishi $\&$ Nakanishi (2013)\nocite{Imanishi2013} observed the dense gas tracers HCN/HCO$^{+}$/HNC $J$=4-3  which showed emission that was consistent with starburst-dominated galaxies. Numerical simulations of NGC 1614 suggest a minor merger with the primary galaxy being 3-5 times more massive \citep{Vaisanen2012}.

In this Letter, we present unpublished archival ALMA and new CARMA CO observations for NGC 1614. We combine the new observations with previously published Submillimeter Array (SMA) observations to constrain the physical conditions in the molecular gas using a radiative transfer code and a Bayesian likelihood method. In Section 2, we describe the observations and line ratio maps. In Section 3, we discuss the morphology of NGC 1614 seen in the new observations. In Section 4, we present the results of the radiative transfer analysis, compare them to other LIRGs and discuss the \co-to-\tco\ abundance ratio, \xco. In Section 5, we discuss the CO-to-H$_{2}$ conversion factor, $\alpha_{CO}$ and put limits on the \co-to-H$_{2}$ abundance ratio, \xh.

\section{Observations} 
We use high-resolution observations from CARMA, SMA and ALMA to constrain the physical conditions of the molecular gas (see Table 1 for details). The calibration methods are outlined in the following sections. All datasets have been continuum subtracted using line-free channels and imported into CASA for further processing and imaging. We created datacubes with 20 \kms\ and 100 \kms\ channel widths for \co\ and \tco, respectively. Integrated intensity maps (Figure 1) were created using channels with line emission and all integrated intensity maps have been primary beam corrected. 

\subsection{CARMA}
NGC 1614 was observed using CARMA in the C-array configuration on 2012 December 03 and 18 and D-array configuration on 2012 November 14, 18 and 20 in \co\ and \tco\ $J$=1-0, simultaneously. We use CASA to calibrate and image the $uv$-datasets. The flux calibrator was Uranus and the bandpass and gain calibrator was 0423-013. We combine both array configurations into one dataset for imaging. 

\subsection{SMA}

NGC 1614 was observed with the SMA in \co\ and \tcotwo\ and \cothree\ using the compact array configuration (Wilson et al. 2008). The data processing is described in Wilson et al. (2008). NGC 1614 was also observed with the SMA using the extended and very-extended \citep{Konig2013} array configurations in \co\ and \tco\ $J$=2-1. We obtained the raw data for the extended observations and the calibrated data for the very-extended observations from the SMA archive. We calibrated the extended configuration data using CASA. Uranus, 3C454.3 and 0423-013 were used as the flux, bandpass and gain calibrators, respectively. We combined data from all three configurations to obtain high-resolution maps of \co\ and \tco\ $J$=2-1.  

To recover the short spacings of the SMA \cotwo\ and $J$=3-2 maps, we used the James Clerk Maxwell Telescope (JCMT) to map NGC 1614 in \cotwo\ using the RxA3 receiver on 2012 November 26 and 27 (Program: M12BC14; PI: K. Sliwa) and in \cothree\ using the RxB3 receiver (Wilson et al. 2008). The reduction of the \cothree\ JCMT map is described in Wilson et al. (2008). The \cotwo\ map covered 49\arcsec\ on each side with a 21\arcsec\ beam, ensuring we included all emission from NGC 1614. We created a continuum subtracted datacube with a velocity resolution of 20 \kms\ using the \verb!Starlink! software (Currie et al. 2008)\nocite{Currie2008}. We assume a main beam efficiency ($\eta_{mb}$) of 0.6 for \cothree\ and 0.69 for \cotwo\ to convert the antenna temperature units to main beam temperature ($T_{mb}$). We convert the $T_{mb}$ units to Jy \kms\ using the scaling factors of 26.9 Jy K($T_{mb}$)$^{-1}$ and 22.9 Jy K($T_{mb}$)$^{-1}$ for \cothree\ and $J$=2-1, respectively. We combined the JCMT and SMA data using the ``feathering" method as described in Sliwa et al. (2012)\nocite{Sliwa2012}. 

\subsection{ALMA}

NGC 1614 was observed in \cothree\ and \cosix\ using ALMA during Cycle 0.  We use the calibrated $uv$-datasets obtained from the ALMA archive. We performed two rounds of phase-only self-calibration on both datasets using the \co\ line emission. When imaging the \cothree\ ALMA data, we use the SMA short-spacings recovered \cothree\ datacube as a starting model so as to better constrain the shorter baselines.  The flux scales of the two datasets differ by roughly 10$\%$; however, this is within our assumed calibration uncertainty of 20$\%$. Along with \cothree, \hcofour\ was observed simultaneously in the upper sideband. We present the integrated intensity map in Figure 1, but do not include the line in our analysis. We also create three 435$\mu$m$_{rest}$ continuum maps using the upper sideband (USB), lower sideband (LSB) and both sidebands. Each sideband is roughly 4GHz in width. All three maps are consistent in flux and structure. 

\subsection{Line Ratios}

We created the following line ratio maps: $r_{21}$ = \co\ ($J$=2-1/$J$=1-0),
$r_{32}$ = \co($J$=3-2/$J$=2-1),
$r_{63}$ = \co ($J$=6-5/$J$=3-2),
$^{13}r_{21}$ = \tco ($J$=2-1/$J$=1-0),
$R_{10}$ = \co/\tco\ $J$=1-0,
$R_{21}$ = \co/\tco\ $J$=2-1,
$H_{43}$ = \hcofour/\cothree\ and
$H_{46}$ = \hcofour/\cosix.
We match the angular resolutions of the two maps used to create the individual line ratio maps (Figure 2). We implemented a 3$\sigma$ cutoff on each map except \tcotwo, where a 1$\sigma$ cutoff was used. The  $r_{21}$, $r_{32}$ and $r_{63}$ maps show that within the central ring region, the line ratios vary smoothly with no significant changes until the outer regions of the ring where the line ratios drop. As is normal for LIRGs, the $R_{10}$ and $R_{21}$ line ratios are unusually high when compared to normal disk galaxies. 

\section{Morphology}

Our CARMA \coone\ observations do not resolve the ring structure, instead showing a single compact nucleus. The highest resolution map of the OVRO \coone\ observations of Olsson et al. (2010) is at a slightly better resolution and shows an elongated structure; however, our CARMA observations do show a slight extension to the east. Our CARMA observations recover roughly 3 times more flux than the lower resolution map of Olsson et al. (2010), which may be due to a difference in sensitivity and $uv$-coverage.

The SMA and ALMA data resolve the previously observed ring structure within NGC 1614. The SMA+JCMT \cotwo\ and ALMA+SMA \cothree\ maps show extended structure to the east and to the southwest beyond the ring of molecular gas. The extended structure in the high-resolution maps is due to an increase in sensitivity of the maps and also the inclusion of short baselines. Overlaying the high-resolution \cothree\ map on an optical HST image shows that the eastern structure runs along the dust lanes in the southeastern spiral arm, while the southwestern structure extends into the dust features to the west (Figure 1 bottom middle). This suggests that the gas and dust are associated and well mixed together.  

The ALMA \cosix\ observations are the first of their kind for NGC 1614. The ring structure is also seen in \cosix\ and 435 $\mu$m$_{rest}$ continuum; however, the extended features seen in \cotwo\ and $J$=3-2 are not detected. The lack of detection suggests that the molecular gas outside of the ring is more diffuse with physical conditions that favour the low-$J$ CO transitions. In addition, any \cosix\ emission outside the ring will come from diffuse structures that are filtered out at high resolutions with ALMA. 
Finally, we measure a total \cosix\ flux of 1320 $\pm$ 40 Jy \kms\ from the Herschel FTS spectrum of NGC 1614. This shows that the ALMA map is missing roughly 30$\%$ of the \cosix\ flux which may well reside in a diffuse structure.

\section{Molecular Gas Physical Conditions}

To constrain the physical conditions of the molecular gas, we use the radiative transfer code RADEX (van der Tak et al. 2007)\nocite{Vandertak2007}. For \co, we create a grid of temperature (\tkin; range=10$^{0.7}$-10$^{3.8}$ K, 71 points), density (\nhtwo; range=10$^{1.0}$-10$^{7.0}$ cm$^{-3}$, 71 points), column density (\nco; range=10$^{12.0}$-10$^{22.0}$ cm$^{-2}$, 81 points) and filling factor (\ff; range=10$^{-5}$-1, 71 points). For \tco, we create a similar grid with a wide range of  \co-to-\tco\ abundance ratios (\xco; range=10-10$^{5}$, 51 points).

In addition to the RADEX grids, we use a Bayesian likelihood code to create probability distributions for the various parameters. Two calculated values are presented for each parameter: 1DMax and 4DMax. The 1DMax is the maximum likelihood of a specific parameter from the one-dimensional distribution for that parameter. The 4DMax is the maximum likelihood of a single grid point based on the four-dimensional distribution of the parameters that comprise the grid. For more details on the likelihood code, see Kamenetzky et al. (2012)\nocite{Kamenetzky2012}. We also implemented the three priors described in Sliwa et al. (2013)\nocite{Sliwa2013}.

We focus our analysis on the central ring structure because outside the ring the missing flux is seen to be $>$20$\%$, our assumed calibration uncertainty, in our short spacing recovered \cotwo\ and $J$=3-2 maps . We model the physical conditions at a resolution of 0.7\arcsec$\times$0.6\arcsec\ (230 pc $\times$ 200 pc) at the five peak intensity positions (Figure 1) around the ring: 
north ($\alpha_{J2000}$ = 04$^{h}$34$^{m}$00$^{s}$.000, $\delta_{J2000}$ = -08$^{\circ}$34\arcmin44\arcsec.251), 
south ($\alpha_{J2000}$ = 04$^{h}$34$^{m}$00$^{s}$.020, $\delta_{J2000}$ = -08$^{\circ}$34\arcmin45\arcsec.844), 
east ($\alpha_{J2000}$ = 04$^{h}$34$^{m}$00$^{s}$.080, $\delta_{J2000}$ = -08$^{\circ}$34\arcmin45\arcsec.006), 
west ($\alpha_{J2000}$ = 04$^{h}$33$^{m}$59$^{s}$.980, $\delta_{J2000}$ = -08$^{\circ}$34\arcmin45\arcsec.064) 
and the ``hole" ($\alpha_{J2000}$ = 04$^{h}$34$^{m}$00$^{s}$.027, $\delta_{J2000}$ = -08$^{\circ}$34\arcmin45\arcsec.075). Since we do not have high-resolution \co\ and \tcoone\ observations and \tcotwo\ is not detected at all positions in the ring, we adopt the following line ratios to estimate the flux for these three lines at each position of the ring: $r_{21}$ = 0.93 $\pm$ 0.23, $^{13}r_{21}$ = 2.0 $\pm$ 0.6, and $R_{21}$ = 20 $\pm$ 3. These line ratios, within uncertainty, span the observed values and the line ratios across the ring do not vary dramatically. 

Over the ring the results are quite consistent and reveal a cold (\tkin=20-40K), dense (\nhtwo\ $>$ 10$^{3.0}$ cm$^{-3}$) molecular gas component (Table 2, Figure 3). Interestingly, only the ``hole" shows evidence of a second warmer molecular gas component; it is not possible to fit all four \co\ lines with a single component (Figure 3). We re-model the molecular gas of the ``hole" excluding the \cosix\ flux. The lower-$J$ lines are better fit without \cosix\ revealing again, cold, dense molecular gas. More higher-$J$ CO observations would be required to constrain the properties of the warmer component in NGC 1614. For the other regions, the cold molecular gas component is dominant but it does not rule out a second warmer component as seen with the Antennae (Schirm et al. 2014), Arp 220 (Rangwala et al. 2011) and M82 (Kamenetzky et al. 2012).

The molecular gas in Arp 299 (Sliwa et al. 2012) and the Antennae (Schirm et al. 2014) is similar to the molecular gas in the ring which could suggest that NGC 1614 is at a similar merger stage to these two LIRGs. This is consistent with age estimates using numerical modeling, which suggest that NGC1614 is $\sim$50 Myr after the second passage \citep{Vaisanen2012} and the Antennae is at a similar stage (Privon et al. 2013)\nocite{Privon2013}. In contrast, the molecular gas of the late stage mergers VV114 (Sliwa et al. 2013) and Arp220 (Matsushita et al. 2009)\nocite{Matsushita2009} is more diffuse ($<$ 10$^{3.0}$ cm$^{-3}$) and warmer ($<$30K), respectively. Note that all these analyses differ in their physical resolutions.

Sliwa et al. (2013) showed that the \xco\ ratio for the LIRG VV 114 is roughly 3 larger than in the local ISM. For NGC 1614, we see similar results for \xco, where the maximum likelihood value exceeds 130 around the ring. The ``hole" has a much lower most probable \xco\ value, similar to the value for the inner regions of the Galaxy (Langer$\&$ Penzias 1990)\nocite{Langer1990}. In the Galaxy, \xco\ increases with increasing radius. During the merger process, molecular gas can be driven towards the central regions, which can bring in high \xco\ valued molecular gas and drive up the \xco\ ratio (Casoli et al. 1992)\nocite{Casoli1992}. Since the ``hole" has a maximum likelihood \xco\ value much lower than the ring, this implies that the infalling molecular gas has not reached the ``hole" and supports the idea of a resonance phenomenon slowing down the infall of molecular gas at certain radii \citep{Konig2013}.

\section{CO-to-H$_{2}$ Conversion Factor}

Narayanan et al. (2011)\nocite{Narayanan2011} found that the combination of increased velocity dispersion and \tkin\ increases the CO intensity, and thus lowers the \co\ luminosity to H$_{2}$ mass conversion factor, $\alpha_{CO}$ (=M$_{H_{2}}$/L$_{CO}$), by $\sim$2-10 times from the Galactic value ($\alpha_{CO;Galactic}$ = 4.3 \alphaco; Bolatto, Wolfire $\&$ Leory 2013)\nocite{Bolatto2013} in mergers. 

We measure $\alpha_{CO}$ using the following
\begin{equation}
\alpha_{CO} = \frac{N_{^{12}CO}\cdot\textrm{area}\cdot m_{H_{2}}}{M_{\odot}\cdot x_{CO}} \frac{r_{21}}{L_{CO J=2-1}}
\end{equation}
where \nco\ is the 4DMax column density, area is calculated from the beam size, $m_{H_{2}}$ is the mass of molecular hydrogen, \xh\ is the CO-to-H$_{2}$ abundance ratio assumed to be 3 $\times$ 10$^{-4}$ (Ward et al. 2003)\nocite{Ward2003} and $L_{CO J=2-1}$ is the \cotwo\ luminosity in K \kms\ pc$^{2}$. We use the line ratio $r_{21}$=0.93 $\pm$ 0.23 to estimate the \coone\ luminosity from the \cotwo\ luminosity at $\sim$0.7\arcsec\ (230pc) resolution. Around the ring, $\alpha_{CO}$ varies from 0.9 - 1.5 $(\frac{3 \times 10^{-4}}{x_{\mathrm{co}}}) \textrm{$M_{\odot}$(K km s$^{-1}$ pc$^{2}$)$^{-1}$}$ (Table 2). The ``hole" has a measured $\alpha_{CO}$ that is nearly an order of magnitude lower.

The ring of NGC 1614 has intermediate $\alpha_{CO}$ values when compared to other LIRGs such as Arp 299 (0.4 \alphaco; Sliwa et al. 2012) and VV 114 (0.5 \alphaco; Sliwa et al. 2013). The Antennae was measured to have an $\alpha_{CO}$ of $\sim$7 \alphaco\ on a 390 pc scale (Wilson et al. 2003)\nocite{Wilson2003}. The difference in $\alpha_{CO}$ value between the Antennae and NGC 1614 could suggest two scenarios: the transition from a Galactic-like value to a ULIRG-like value (Downes $\&$ Solomon 1998)\nocite{Downes1998} is extremely rapid or the Antennae has yet to hit its peak star formation rate. Narayanan et al. (2011) showed that during the merger process, $\alpha_{CO}$ drops to its lowest value around the star formation peak. The star formation rate of the Antennae is nearly an order of magnitude lower than that of NGC 1614. 

The biggest uncertainty in measuring $\alpha_{CO}$ is the assumed value of \xh. Clouds in Taurus and $\rho$ Ophiuchi have a measured value of \xh\ = 8.5 $\times$ 10$^{-5}$ (Frerking, Langer $\&$ Wilson 1982)\nocite{Frerking1982}, while the ratio measured in warm, star forming clouds can be higher, such as NGC 2024 with a measured \xh\ of 2.7 $\times$ 10$^{-4}$ (Lacy et al. 1994)\nocite{Lacy1994}. Since NGC 1614 is undergoing a starburst (SFR $\sim$ 52 \msol\ yr$^{-1}$; U et al. 2012)\nocite{U2012}, we assume a high \xh\ value, similar to that of NGC 2024. The ``hole" in the molecular ring, however, is not forming stars vigorously and the molecular gas is colder; therefore, the \xh\ that we have adopted may not be appropriate for the ``hole". If \xh\ is lower in the ``hole", this would drive up the measured $\alpha_{CO}$ value. 

Schirm et al. (2014)\nocite{Schirm2014} found an \xh\ of 5 $\times$ 10$^{-5}$ for the Antennae by comparing their hot molecular gas mass to that derived using $Spitzer$ H$_{2}$ vibrational transitions. We are unable to perform such an analysis for NGC 1614 due to the lack of high-$J$ CO observations; however, we can place lower limits on \xh\ using the dynamical masses (\mdyn) of the regions around the ring. We measure \mdyn\ using the line width and the source size, both measured using the \cothree\ map. We then use \mdyn\ and the CO luminosity within the region to measure $\alpha_{CO, dyn}$ (=\mdyn/L$_{CO}$). Since \mdyn\ is the mass of all matter (i.e. stars, gas, dust, and dark matter), $\alpha_{CO, dyn}$ is an upper limit,
\begin{equation}
\alpha_{CO, dyn} \geq \alpha_{CO, true} \propto \frac{1}{x_{CO}}
\end{equation}
In addition, the measured \mdyn\ is only valid for regions that are gravitationally bound; however, given their size, our regions are likely not gravitationally bound and so the measured \mdyn\ is itself also an upper limit. For the east, west and south regions, we measure $\alpha_{CO,dyn}$ to be 3.5, 2.6 and 2.4 \alphaco, respectively. This sets a lower limit on \xh\ of 1.1-1.4 $\times$ 10$^{-4}$. For the north region, we get an interesting result: $\alpha_{CO, dyn}$ is exactly what we measured from the radiative transfer analysis. This implies that our assumption of \xh\ for the north region is likely not valid and should be higher by some factor which would drive $\alpha_{CO}$ to lower values. If we assume that the ratio between the \mdyn\ and radiative transfer $\alpha_{CO}$  is similar for all regions around the ring (excluding the ``hole"), \xh\ should be $\sim$2-3 times larger than our assumption of 3 $\times$ 10$^{-4}$ driving the value from 1.2 \alphaco\ to 0.4-0.6 \alphaco.

One thing is certain about $\alpha_{CO}$ for NGC 1614: the Galactic value does not apply and a more ULIRG-type value should be used to measure the amount of molecular gas. With future ALMA observations of other high-$J$ CO lines and James Webb Space Telescope observations of vibrational H$_{2}$ transitions in the near-infrared, we will be able to put better constraints on \xh\ and in turn $\alpha_{CO}$ for NGC 1614.

\acknowledgments{We thank the referee for his/her comments. The SMA is a joint project between the Smithsonian Astrophysical Observatory and the Academia Sinica Institute of Astronomy and Astrophysics and is funded by the Smithsonian Institution and the Academia Sinica. 
The JCMT has historically been operated by the JAC on behalf of the STFC of the United Kingdom, the NRC of Canada and the Netherlands Organisation for Scientific Research.
Support for CARMA construction was derived from the states of California, Illinois, and Maryland, the James S. McDonnell Foundation, the Gordon and Betty Moore Foundation, the Kenneth T. and Eileen L. Norris Foundation, the University of Chicago, the Caltech Associates, and the NSF. Ongoing CARMA development and operations are supported by the NSF under a cooperative agreement, and by the CARMA partner universities. This paper makes use of the following ALMA data: ADS/JAO.ALMA$\#$2011.0.00182.S and ADS/JAO.ALMA$\#$2011.0.00768.S. ALMA is a partnership of ESO (representing its member states), NSF (USA) and NINS (Japan), together with NRC (Canada) and NSC and ASIAA (Taiwan), in cooperation with the Republic of Chile. The Joint ALMA Observatory is operated by ESO, AUI/NRAO and NAOJ. The NRAO is a facility of the NSF operated under cooperative agreement by Associated Universities, Inc. The research of C.D.W. is supported by NSERC Canada. The research of K.S. is supported by OGS. }

{\it Facilities:} \facility{ALMA, CARMA, SMA, JCMT}.


\begin{thebibliography}{28}
\expandafter\ifx\csname natexlab\endcsname\relax\def\natexlab#1{#1}\fi

\bibitem[{{Alonso-Herrero} {et~al.}(2001){Alonso-Herrero}, {Engelbracht},
  {Rieke}, {Rieke}, \& {Quillen}}]{AlonsoHerrero2001}
{Alonso-Herrero}, A., {Engelbracht}, C.~W., {Rieke}, M.~J., {Rieke}, G.~H., \&
  {Quillen}, A.~C. 2001, \apj, 546, 952

\bibitem[{{Bolatto} {et~al.}(2013){Bolatto}, {Wolfire}, \&
  {Leroy}}]{Bolatto2013}
{Bolatto}, A.~D., {Wolfire}, M., \& {Leroy}, A.~K. 2013, \araa, 51, 207

\bibitem[{{Casoli} {et~al.}(1992){Casoli}, {Dupraz}, \& {Combes}}]{Casoli1992}
{Casoli}, F., {Dupraz}, C., \& {Combes}, F. 1992, \aap, 264, 55

\bibitem[{{Currie} {et~al.}(2008){Currie}, {Draper}, {Berry}, {Jenness},
  {Cavanagh}, {et~al.}}]{Currie2008}
{Currie}, M.~J., {Draper}, P.~W., {Berry}, D.~S., {Jenness}, T., {Cavanagh},
  B., {et~al.} 2008, in Astronomical Society of the Pacific Conference Series,
  Vol. 394, Astronomical Data Analysis Software and Systems XVII, ed. R.~W.
  {Argyle}, P.~S. {Bunclark}, \& J.~R. {Lewis}, 650

\bibitem[{{Downes} \& {Solomon}(1998)}]{Downes1998}
{Downes}, D. \& {Solomon}, P.~M. 1998, \apj, 507, 615

\bibitem[{{Frerking} {et~al.}(1982){Frerking}, {Langer}, \&
  {Wilson}}]{Frerking1982}
{Frerking}, M.~A., {Langer}, W.~D., \& {Wilson}, R.~W. 1982, \apj, 262, 590

\bibitem[{{Imanishi} \& {Nakanishi}(2013)}]{Imanishi2013}
{Imanishi}, M. \& {Nakanishi}, K. 2013, \aj, 146, 47

\bibitem[{{Kamenetzky} {et~al.}(2012){Kamenetzky}, {Glenn}, {Rangwala},
  {Maloney}, {Bradford}, {et~al.}}]{Kamenetzky2012}
{Kamenetzky}, J., {Glenn}, J., {Rangwala}, N., {Maloney}, P., {Bradford}, M.,
  {et~al.} 2012, \apj, 753, 70

\bibitem[{{K{\"o}nig} {et~al.}(2013){K{\"o}nig}, {Aalto}, {Muller}, {Beswick},
  \& {Gallagher}}]{Konig2013}
{K{\"o}nig}, S., {Aalto}, S., {Muller}, S., {Beswick}, R.~J., \& {Gallagher},
  J.~S. 2013, \aap, 553, A72

\bibitem[{{Lacy} {et~al.}(1994){Lacy}, {Knacke}, {Geballe}, \&
  {Tokunaga}}]{Lacy1994}
{Lacy}, J.~H., {Knacke}, R., {Geballe}, T.~R., \& {Tokunaga}, A.~T. 1994,
  \apjl, 428, L69

\bibitem[{{Langer} \& {Penzias}(1990)}]{Langer1990}
{Langer}, W.~D. \& {Penzias}, A.~A. 1990, \apj, 357, 477

\bibitem[{{Matsushita} {et~al.}(2009){Matsushita}, {Iono}, {Petitpas}, {Chou},
  {Gurwell}, {et~al.}}]{Matsushita2009}
{Matsushita}, S., {Iono}, D., {Petitpas}, G.~R., {Chou}, R.~C.-Y., {Gurwell},
  M.~A., {et~al.} 2009, \apj, 693, 56

\bibitem[{{Mihos} \& {Hernquist}(1994)}]{Mihos1994}
{Mihos}, J.~C. \& {Hernquist}, L. 1994, \apjl, 425, L13

\bibitem[{{Mihos} \& {Hernquist}(1996)}]{Mihos1996}
---. 1996, \apj, 464, 641

\bibitem[{{Narayanan} {et~al.}(2011){Narayanan}, {Krumholz}, {Ostriker}, \&
  {Hernquist}}]{Narayanan2011}
{Narayanan}, D., {Krumholz}, M., {Ostriker}, E.~C., \& {Hernquist}, L. 2011,
  \mnras, 418, 664

\bibitem[{{Neff} {et~al.}(1990){Neff}, {Hutchings}, {Standord}, \&
  {Unger}}]{Neff1990}
{Neff}, S.~G., {Hutchings}, J.~B., {Standord}, S.~A., \& {Unger}, S.~W. 1990,
  \aj, 99, 1088

\bibitem[{{Olsson} {et~al.}(2010){Olsson}, {Aalto}, {Thomasson}, \&
  {Beswick}}]{Olsson2010}
{Olsson}, E., {Aalto}, S., {Thomasson}, M., \& {Beswick}, R. 2010, \aap, 513,
  A11

\bibitem[{{Privon} {et~al.}(2013){Privon}, {Barnes}, {Evans}, {Hibbard}, {Yun},
  {et~al.}}]{Privon2013}
{Privon}, G.~C., {Barnes}, J.~E., {Evans}, A.~S., {Hibbard}, J.~E., {Yun},
  M.~S., {et~al.} 2013, \apj, 771, 120

\bibitem[{{Sanders} {et~al.}(2003){Sanders}, {Mazzarella}, {Kim}, {Surace}, \&
  {Soifer}}]{Sanders2003}
{Sanders}, D.~B., {Mazzarella}, J.~M., {Kim}, D.-C., {Surace}, J.~A., \&
  {Soifer}, B.~T. 2003, \aj, 126, 1607

\bibitem[{{Schirm} {et~al.}(2014){Schirm}, {Wilson}, {Parkin}, {Kamenetzky},
  {Glenn}, {et~al.}}]{Schirm2014}
{Schirm}, M.~R.~P., {Wilson}, C.~D., {Parkin}, T.~J., {Kamenetzky}, J.,
  {Glenn}, J., {et~al.} 2014, \apj, 781, 101

\bibitem[{{Sliwa} {et~al.}(2013){Sliwa}, {Wilson}, {Krips}, {Petitpas}, {Iono},
  {et~al.}}]{Sliwa2013}
{Sliwa}, K., {Wilson}, C.~D., {Krips}, M., {Petitpas}, G.~R., {Iono}, D.,
  {et~al.} 2013, \apj, 777, 126

\bibitem[{{Sliwa} {et~al.}(2012){Sliwa}, {Wilson}, {Petitpas}, {Armus},
  {Juvela}, {et~al.}}]{Sliwa2012}
{Sliwa}, K., {Wilson}, C.~D., {Petitpas}, G.~R., {Armus}, L., {Juvela}, M.,
  {et~al.} 2012, \apj, 753, 46

\bibitem[{{U} {et~al.}(2012){U}, {Sanders}, {Mazzarella}, {Evans}, {Howell},
  {et~al.}}]{U2012}
{U}, V., {Sanders}, D.~B., {Mazzarella}, J.~M., {Evans}, A.~S., {Howell},
  J.~H., {et~al.} 2012, \apjs, 203, 9

\bibitem[{{V{\"a}is{\"a}nen} {et~al.}(2012){V{\"a}is{\"a}nen}, {Rajpaul},
  {Zijlstra}, {Reunanen}, \& {Kotilainen}}]{Vaisanen2012}
{V{\"a}is{\"a}nen}, P., {Rajpaul}, V., {Zijlstra}, A.~A., {Reunanen}, J., \&
  {Kotilainen}, J. 2012, \mnras, 420, 2209

\bibitem[{{van der Tak} {et~al.}(2007){van der Tak}, {Black}, {Sch{\"o}ier},
  {Jansen}, \& {van Dishoeck}}]{Vandertak2007}
{van der Tak}, F.~F.~S., {Black}, J.~H., {Sch{\"o}ier}, F.~L., {Jansen}, D.~J.,
  \& {van Dishoeck}, E.~F. 2007, \aap, 468, 627

\bibitem[{{Ward} {et~al.}(2003){Ward}, {Zmuidzinas}, {Harris}, \&
  {Isaak}}]{Ward2003}
{Ward}, J.~S., {Zmuidzinas}, J., {Harris}, A.~I., \& {Isaak}, K.~G. 2003, \apj,
  587, 171

\bibitem[{{Wilson} {et~al.}(2008){Wilson}, {Petitpas}, {Iono}, {Baker}, {Peck},
  {et~al.}}]{Wilson2008}
{Wilson}, C.~D., {Petitpas}, G.~R., {Iono}, D., {Baker}, A.~J., {Peck}, A.~B.,
  {et~al.} 2008, \apjs, 178, 189

\bibitem[{{Wilson} {et~al.}(2003){Wilson}, {Scoville}, {Madden}, \&
  {Charmandaris}}]{Wilson2003}
{Wilson}, C.~D., {Scoville}, N., {Madden}, S.~C., \& {Charmandaris}, V. 2003,
  \apj, 599, 1049

\end{thebibliography}


\begin{figure}[h] 
\centering
$\begin{array}{@{\hspace{-0.3in}}c@{\hspace{0.1in}}c@{\hspace{0.1in}}c}
\includegraphics[ scale=0.25]{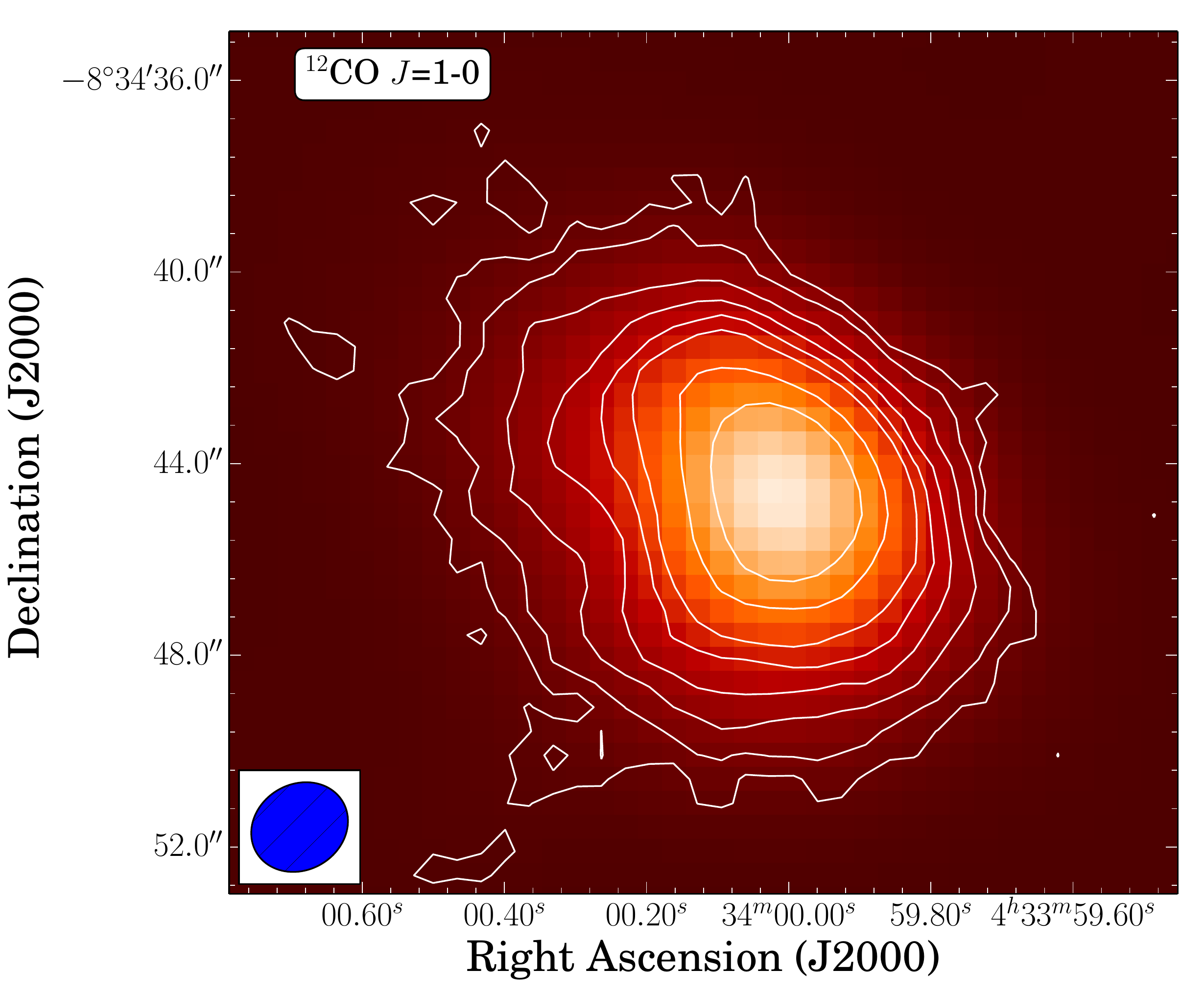}  &\includegraphics[scale=0.25]{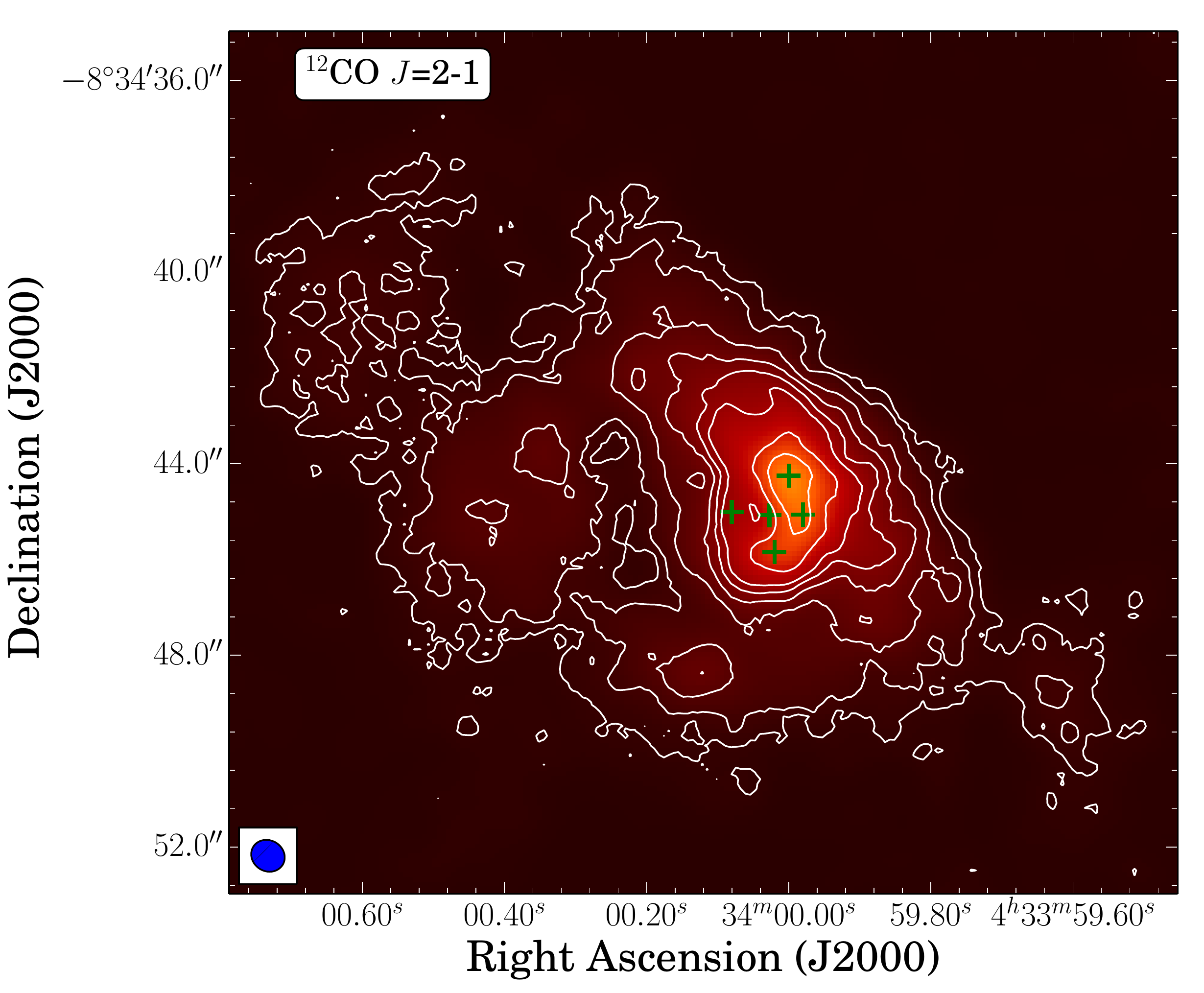} & \includegraphics[scale=0.25]{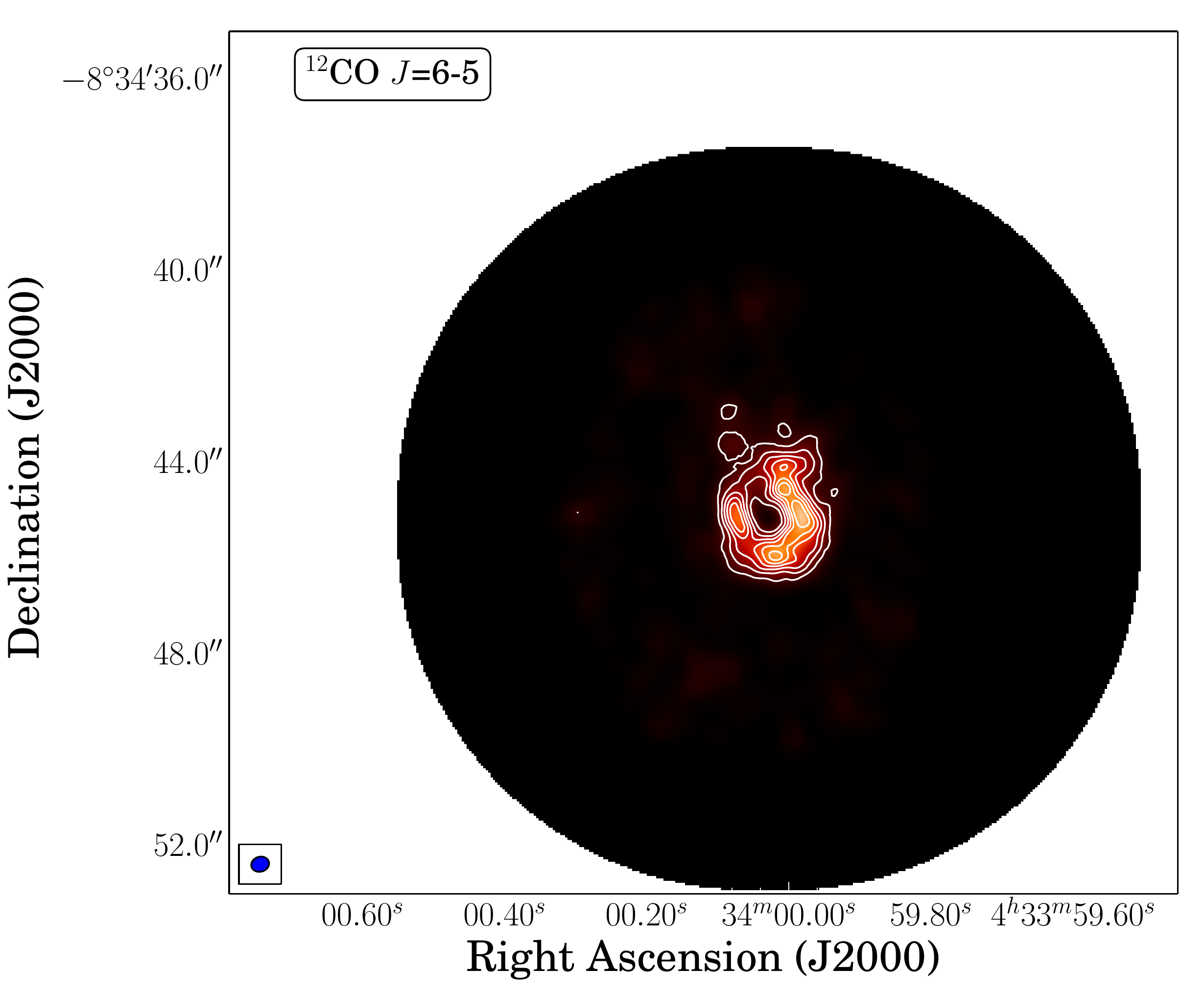} \\
\includegraphics[scale=0.25]{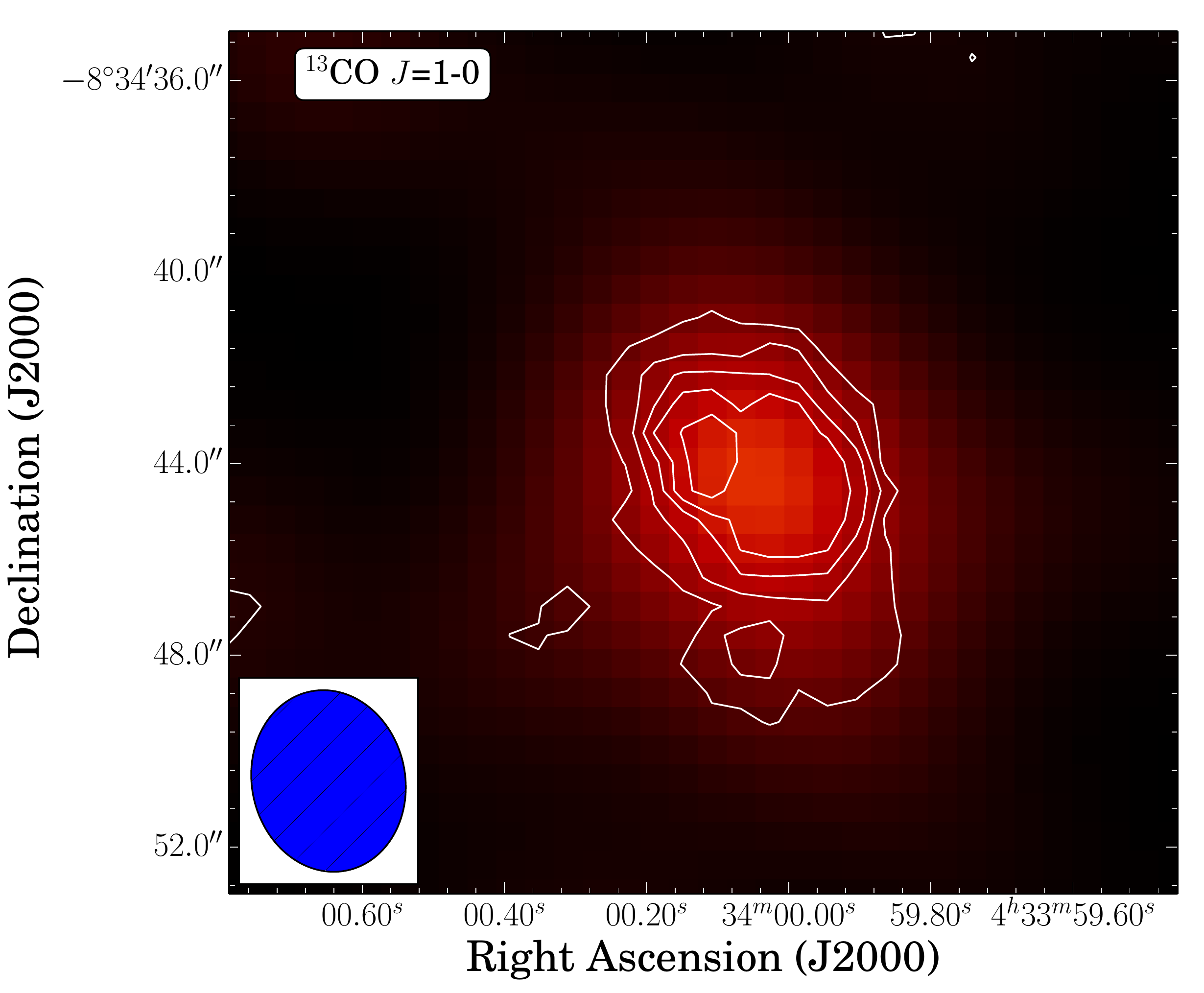}  & \includegraphics[scale=0.25]{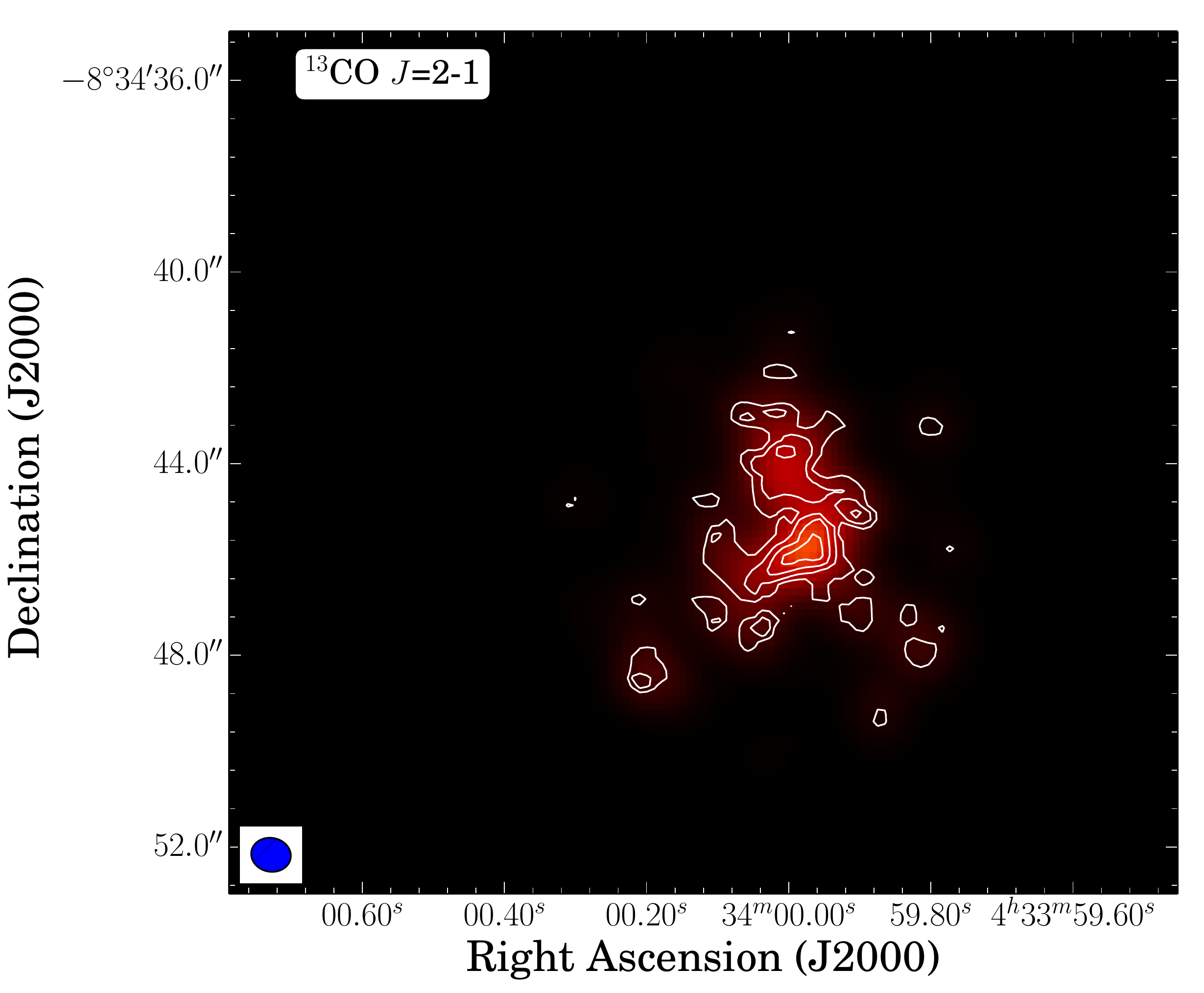} & \includegraphics[scale=0.25]{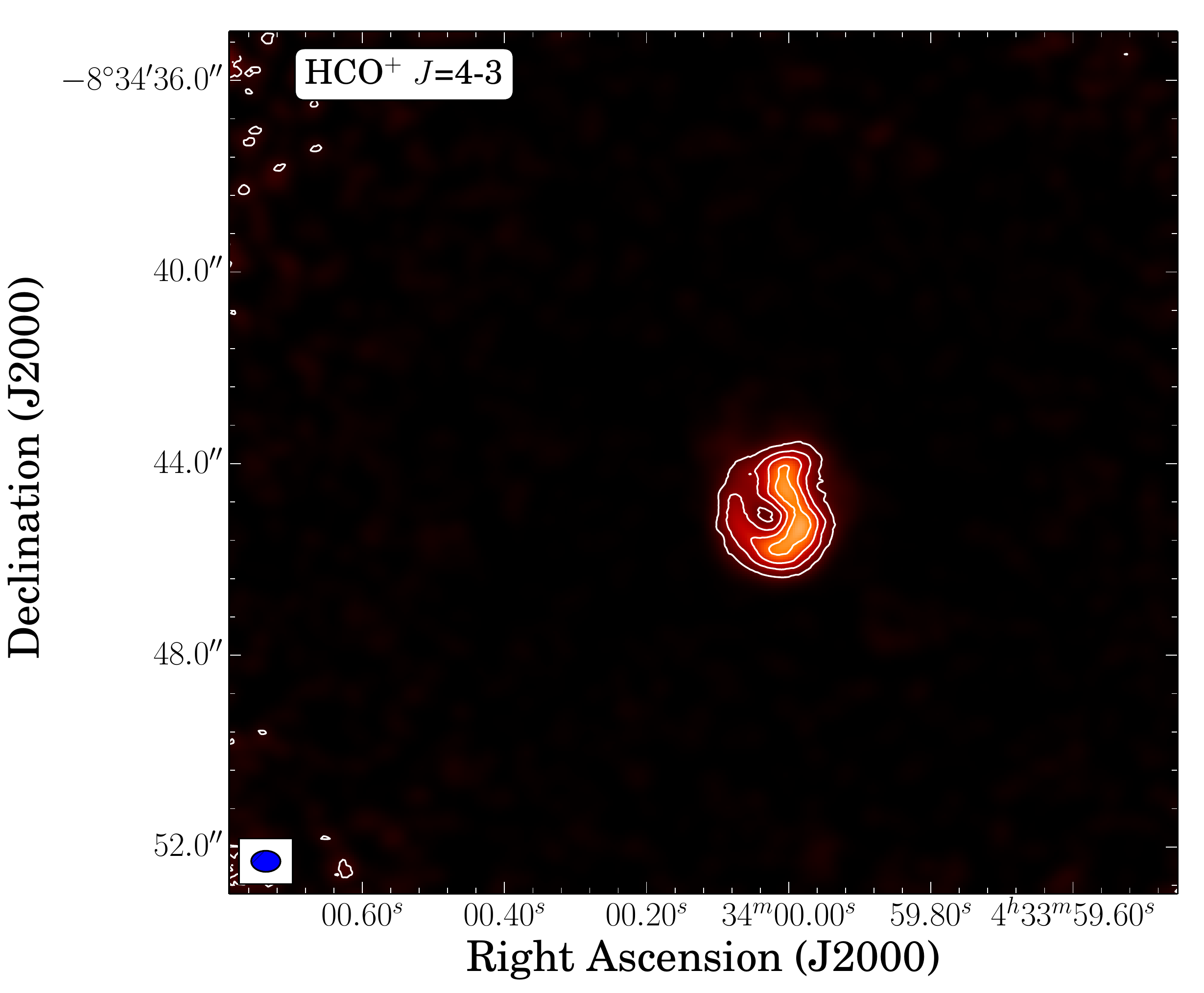} \\ 

\includegraphics[scale=0.25]{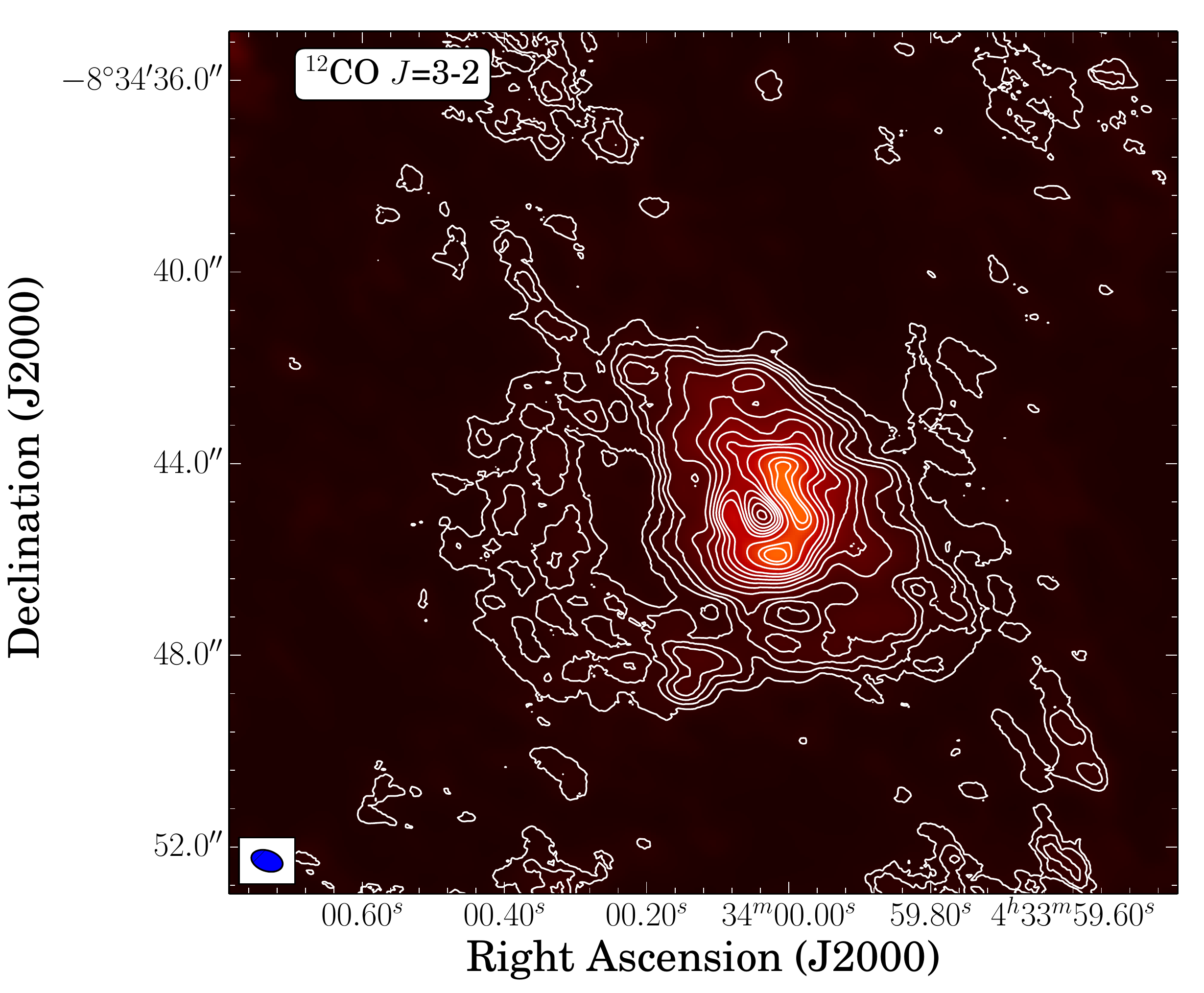} & \includegraphics[scale=0.25]{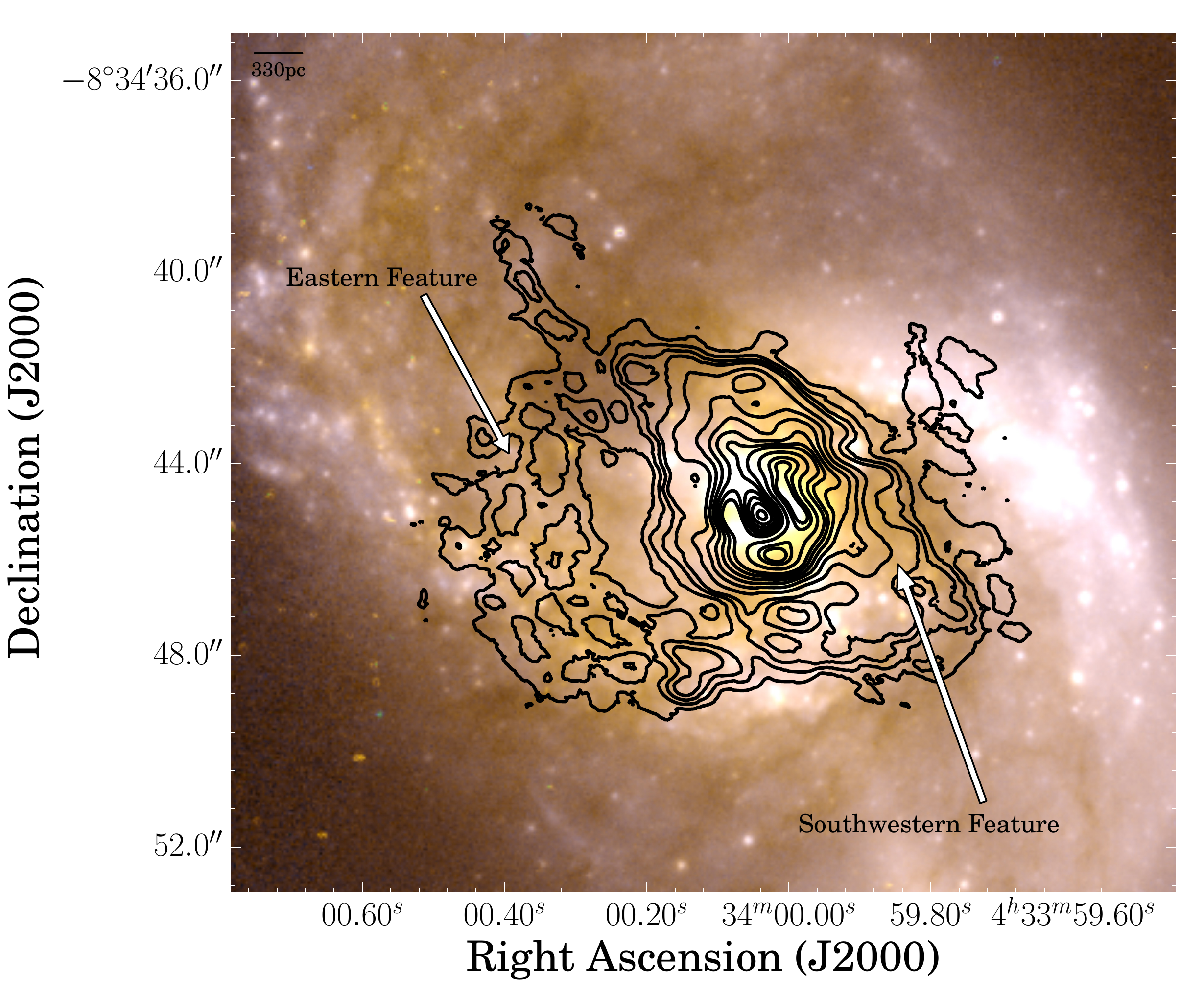} & \includegraphics[scale=0.25]{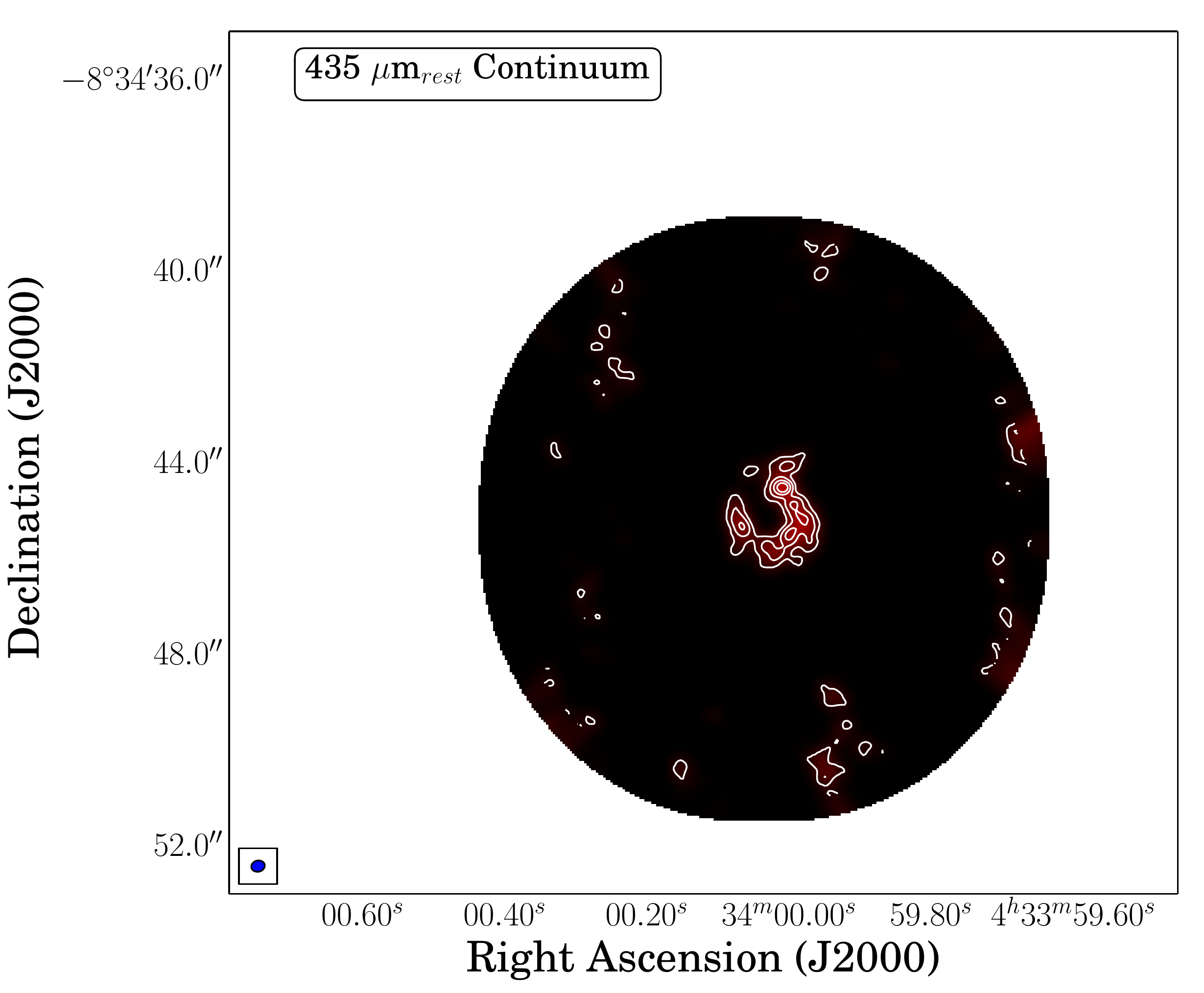}   \\
\end{array}$

\caption[]{Integrated intensity maps for NGC 1614. Ellipses in the lower left corner indicate the beam resolution.
\textit{Top Row:} (Left) CARMA \coone\ map. Contours correspond to (1,2,4,6,8,10,15,20) $\times$ 1.03 Jy beam$^{-1}$ \kms. (Middle) SMA+JCMT \cotwo\ map. Green crosses represent the positions of radiative transfer analysis. Contours correspond to (1,2,4,6,8,10,15,20,30)  $\times$ 1.37 Jy beam$^{-1}$ \kms. (Right) ALMA \cosix\ map. Contours correspond to (5,10,15,20,25,30)  $\times$ 1.26 Jy beam$^{-1}$ \kms.
\textit{Middle Row:} (Left) CARMA \tcoone\ map. Contours correspond to (1,1.5,2,2.5,3) $\times$ 0.62 Jy beam$^{-1}$ \kms.  (Middle) SMA \tcotwo\ map. Contours correspond to (0.5,1,1.5,2,2.5,3)  $\times$ 1.0 Jy beam$^{-1}$ \kms. (Right) ALMA \hcofour\ map. Contours correspond to (3,6,9,12,15)  $\times$ 0.168 Jy beam$^{-1}$ \kms.
\textit{Bottom Row:} (Left) ALMA+SMA \cothree\ map. Contours correspond to (4,6,8,10,15,20,25,30,35,40,50,60,65)  $\times$ 0.64 Jy beam$^{-1}$ \kms. (Middle) ALMA+SMA \cothree\ contours overlaid on HST 814/435 color map. (Right): ALMA 435 $\mu$m$_{rest}$ continuum created with both sidebands. Contours correspond to (3,6,9,12) $\times$ 0.0014 Jy beam$^{-1}$.
  }
\label{SMAmaps}
\end{figure}

\begin{figure}[h] 
\centering
$\begin{array}{@{\hspace{-0.3in}}c@{\hspace{0.05in}}c@{\hspace{0.05in}}c}
\includegraphics[ scale=0.25]{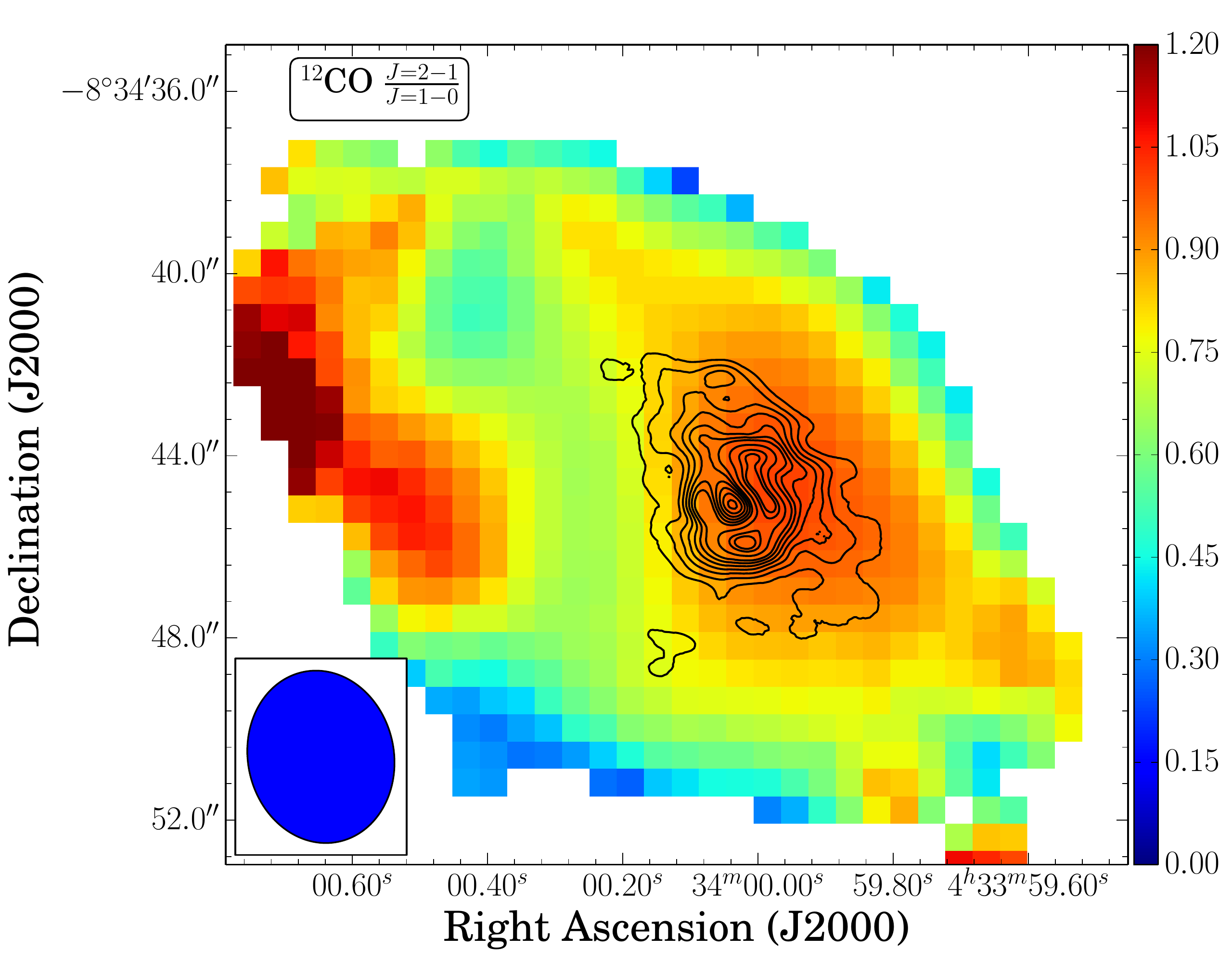}  &\includegraphics[scale=0.25]{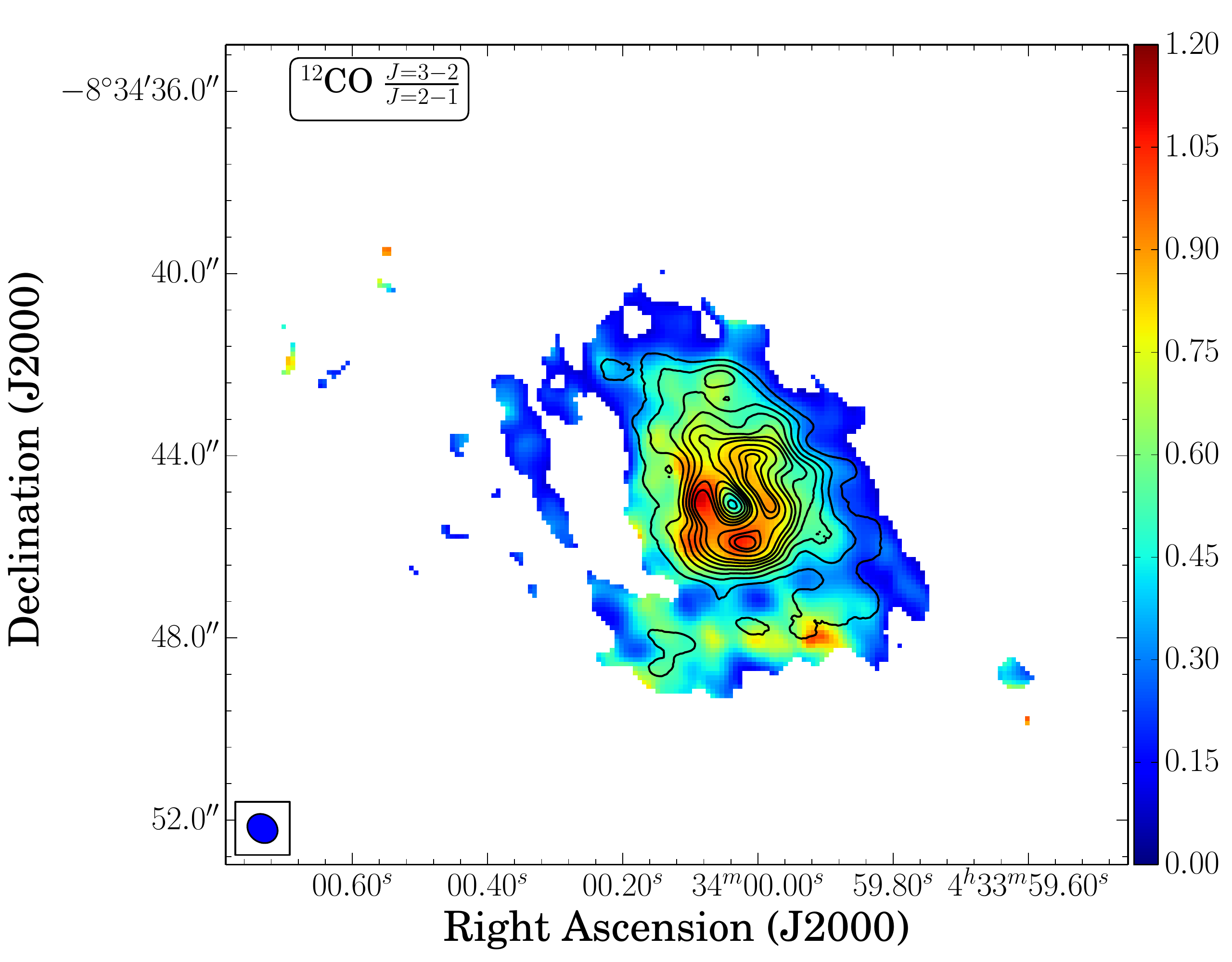} & \includegraphics[scale=0.25]{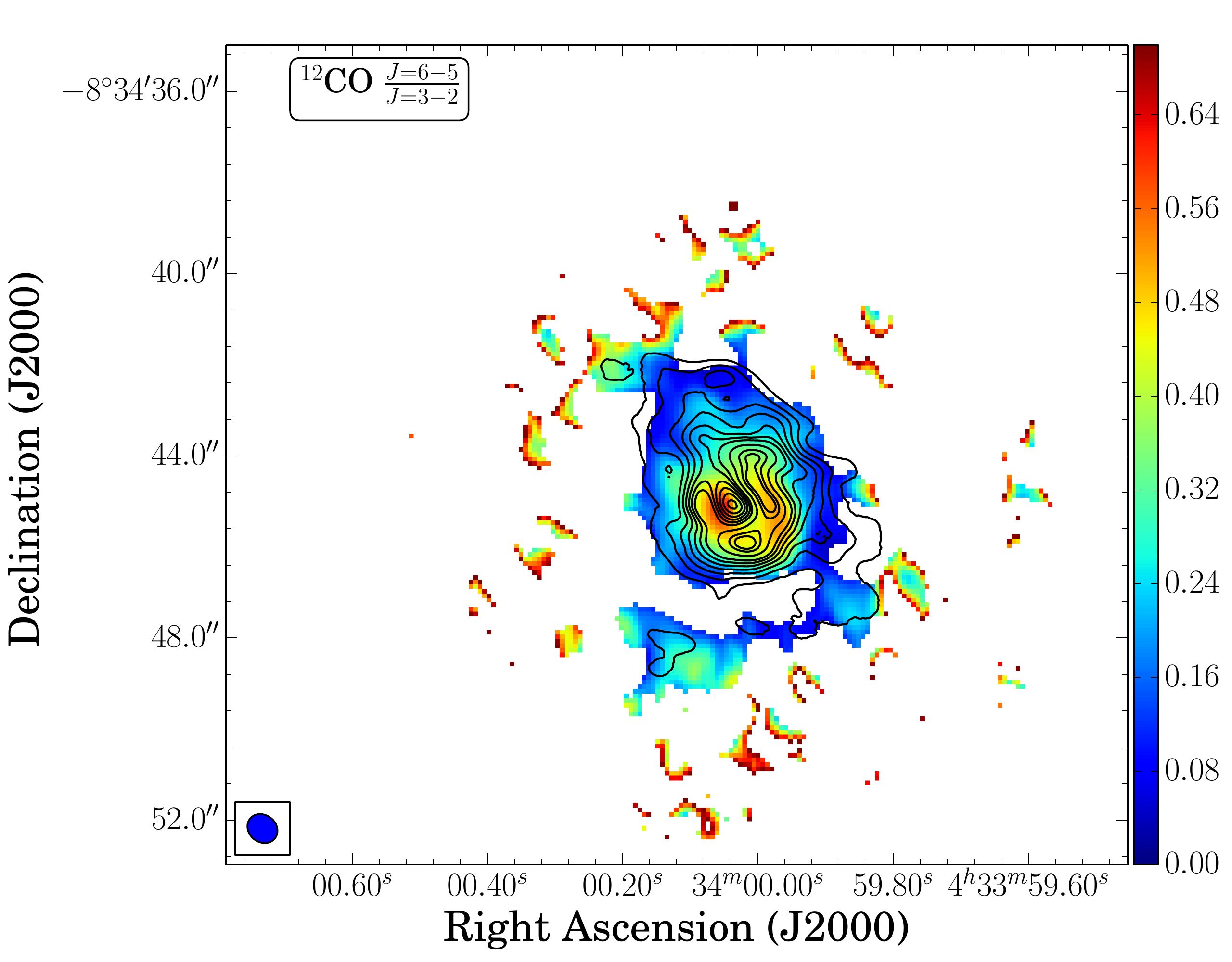} \\
\includegraphics[scale=0.25]{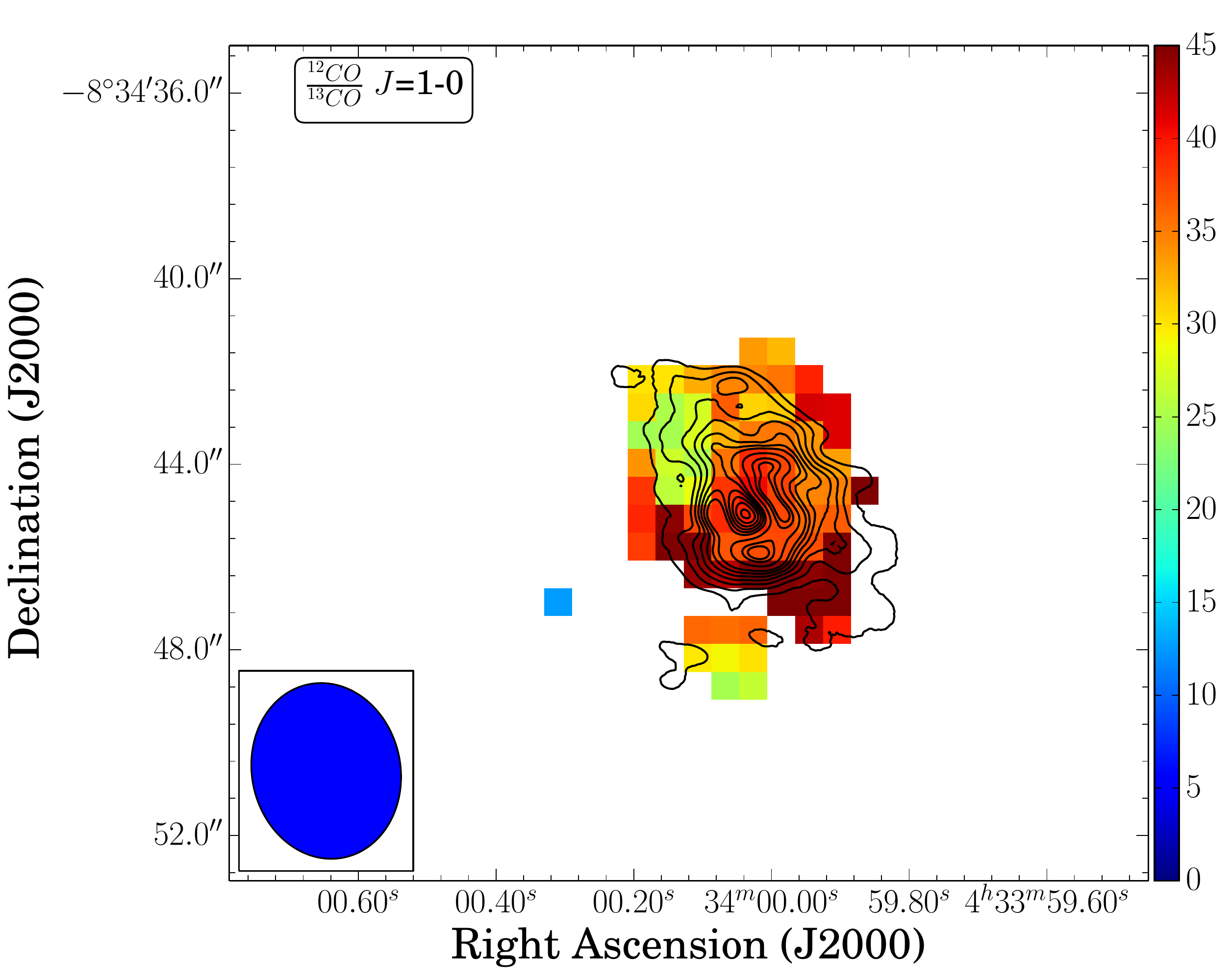}  & \includegraphics[scale=0.25]{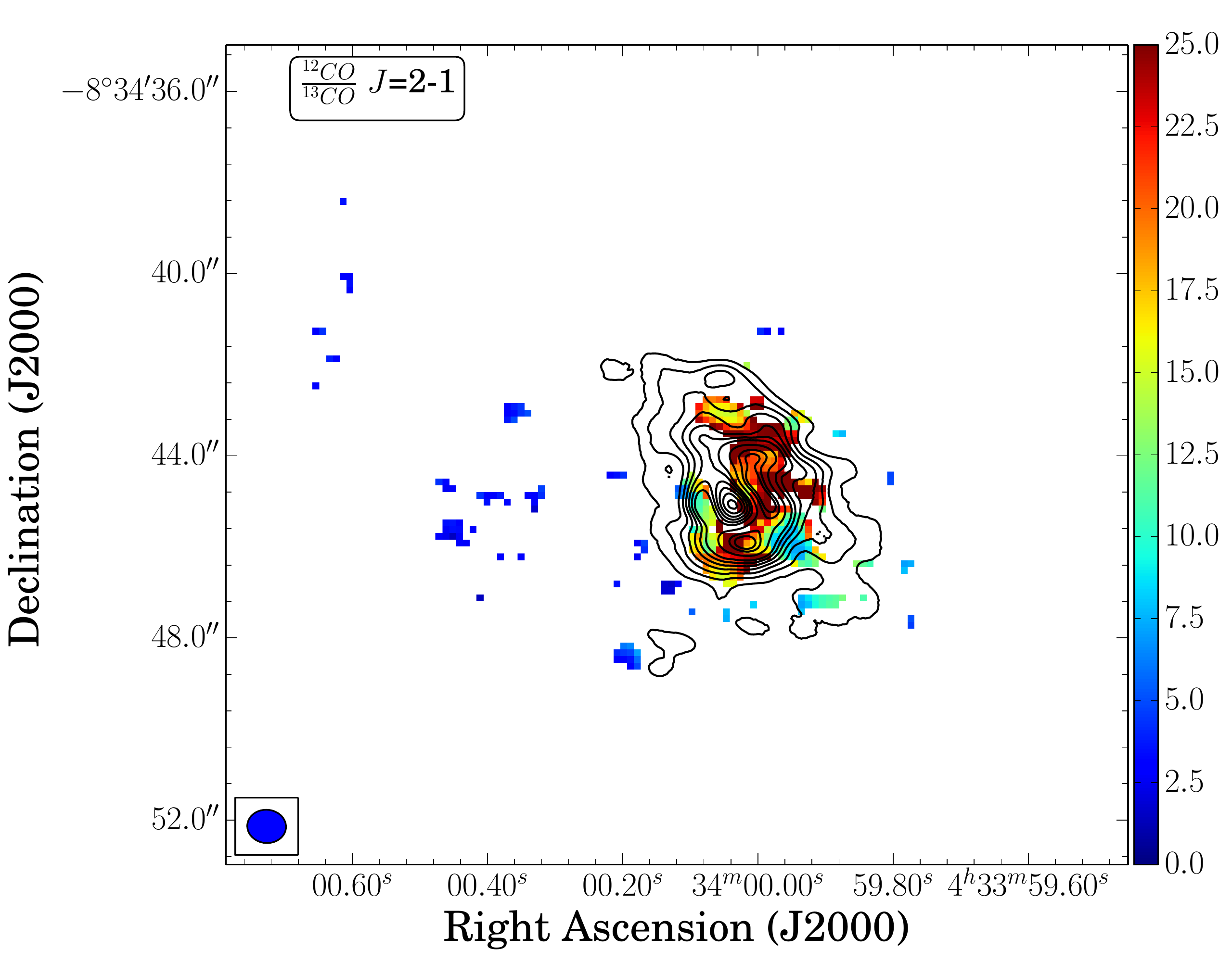} & \includegraphics[scale=0.25]{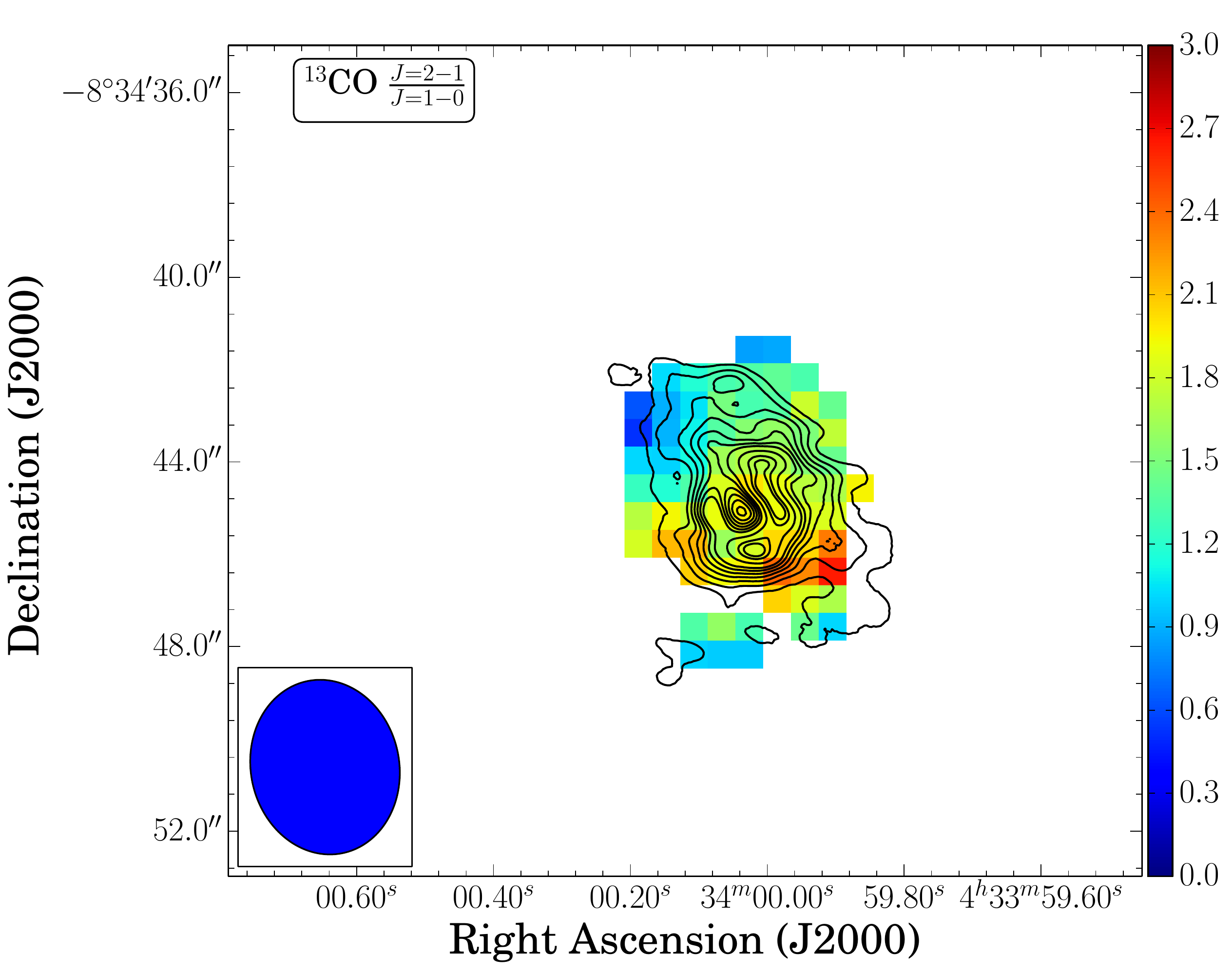} \\ 
\end{array}$

$\begin{array}{@{\hspace{-0.3in}}c@{\hspace{0.1in}}c}
\includegraphics[scale=0.25]{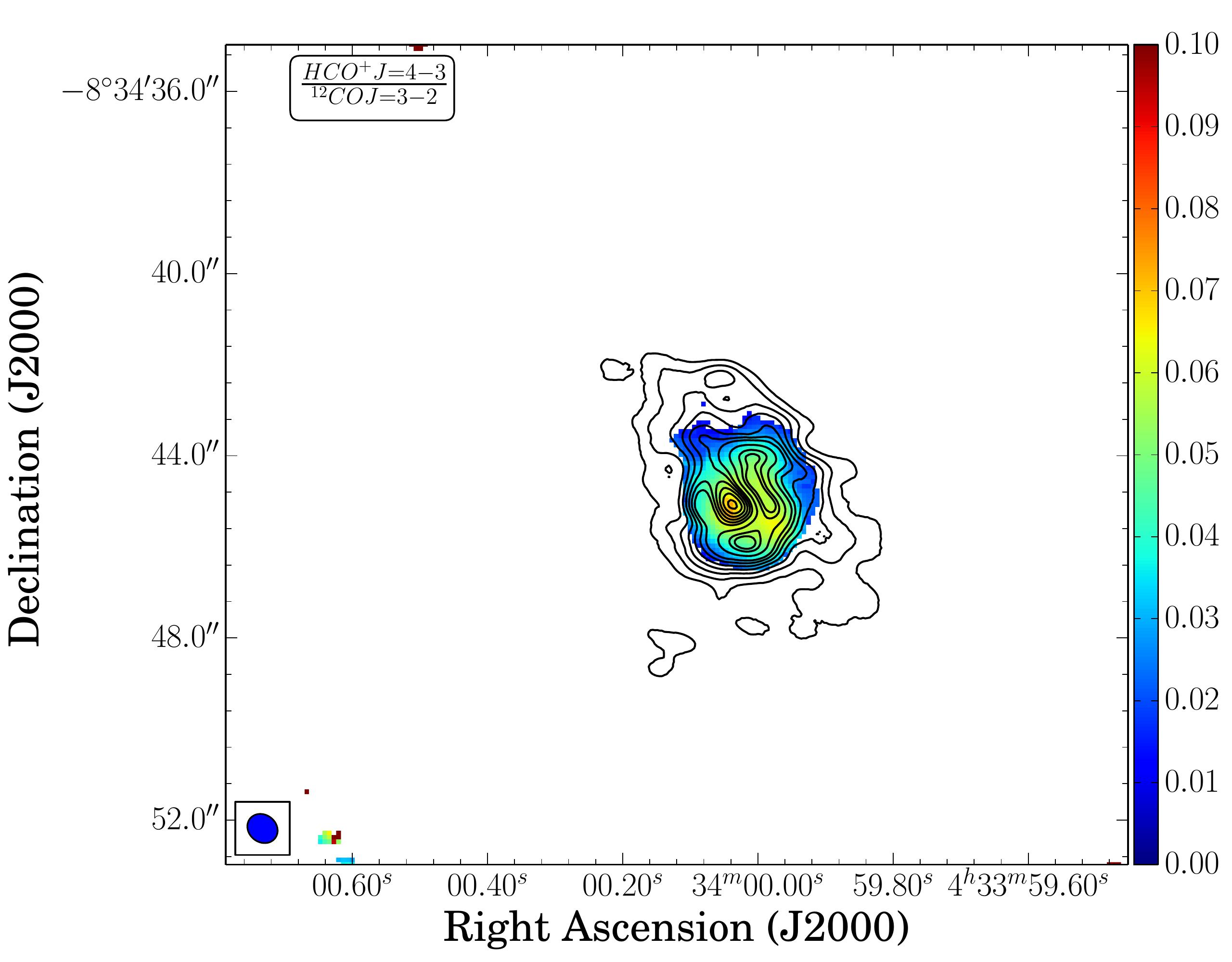} & \includegraphics[scale=0.25]{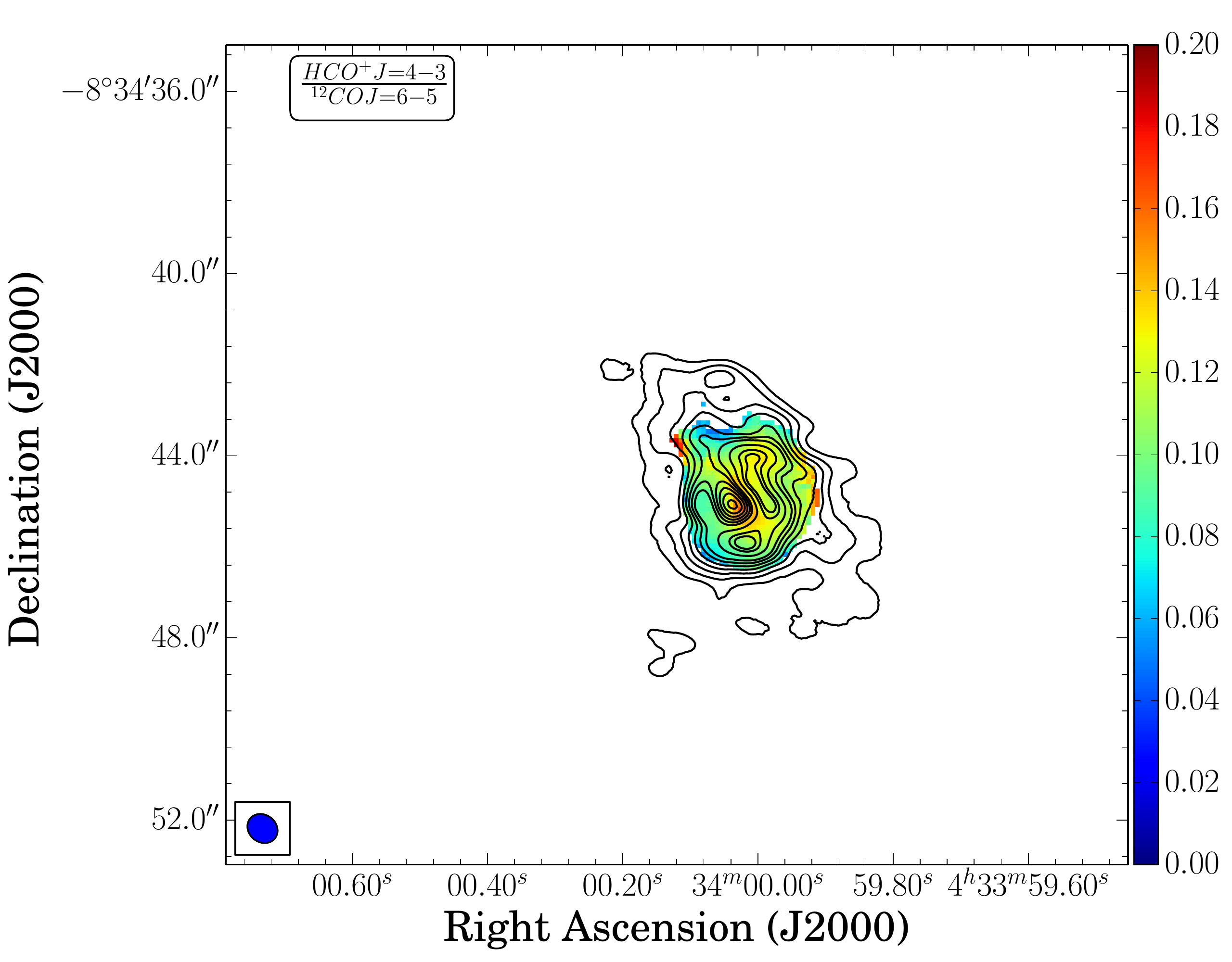}   
\end{array}$

\caption[]{ Line ratio maps. Contours correspond to \cothree\ from Figure 1 starting at 10 $\times$ 0.64 Jy beam$^{-1}$ \kms.
\textit{Top Row:}  $r_{21}$, $r_{32}$ and $r_{63}$
\textit{Middle Row:}  $R_{10}$, $R_{21}$ and $^{13}r_{21}$
\textit{Bottom Row:}  $H_{43}$ and  $H_{46}$. 
}
\label{SMAmaps}
\end{figure}

\begin{figure}[h] 
\centering
$\begin{array}{@{\hspace{-0.3in}}c@{\hspace{0.05in}}c}
\includegraphics[ scale=0.4]{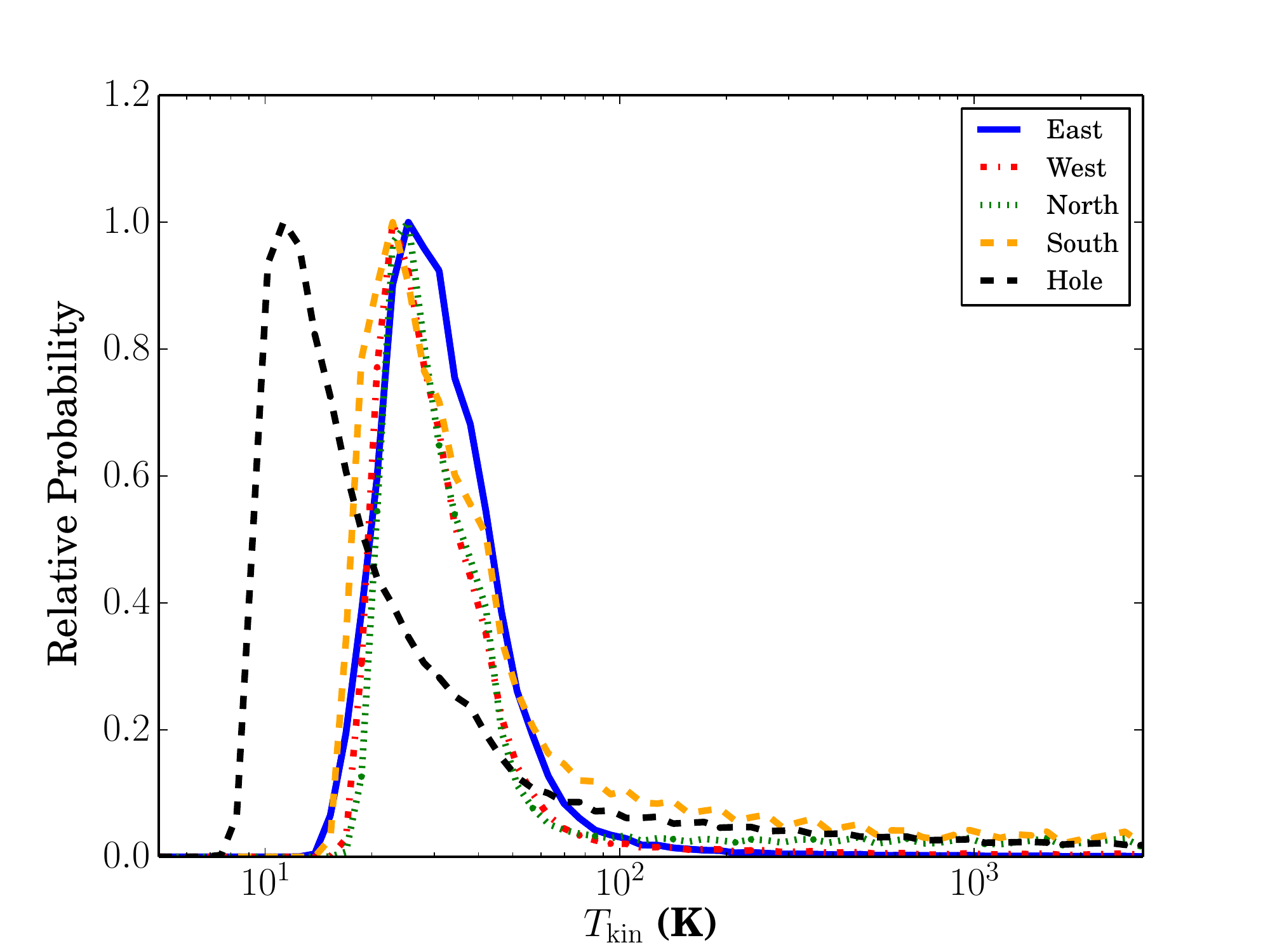}  &\includegraphics[scale=0.4]{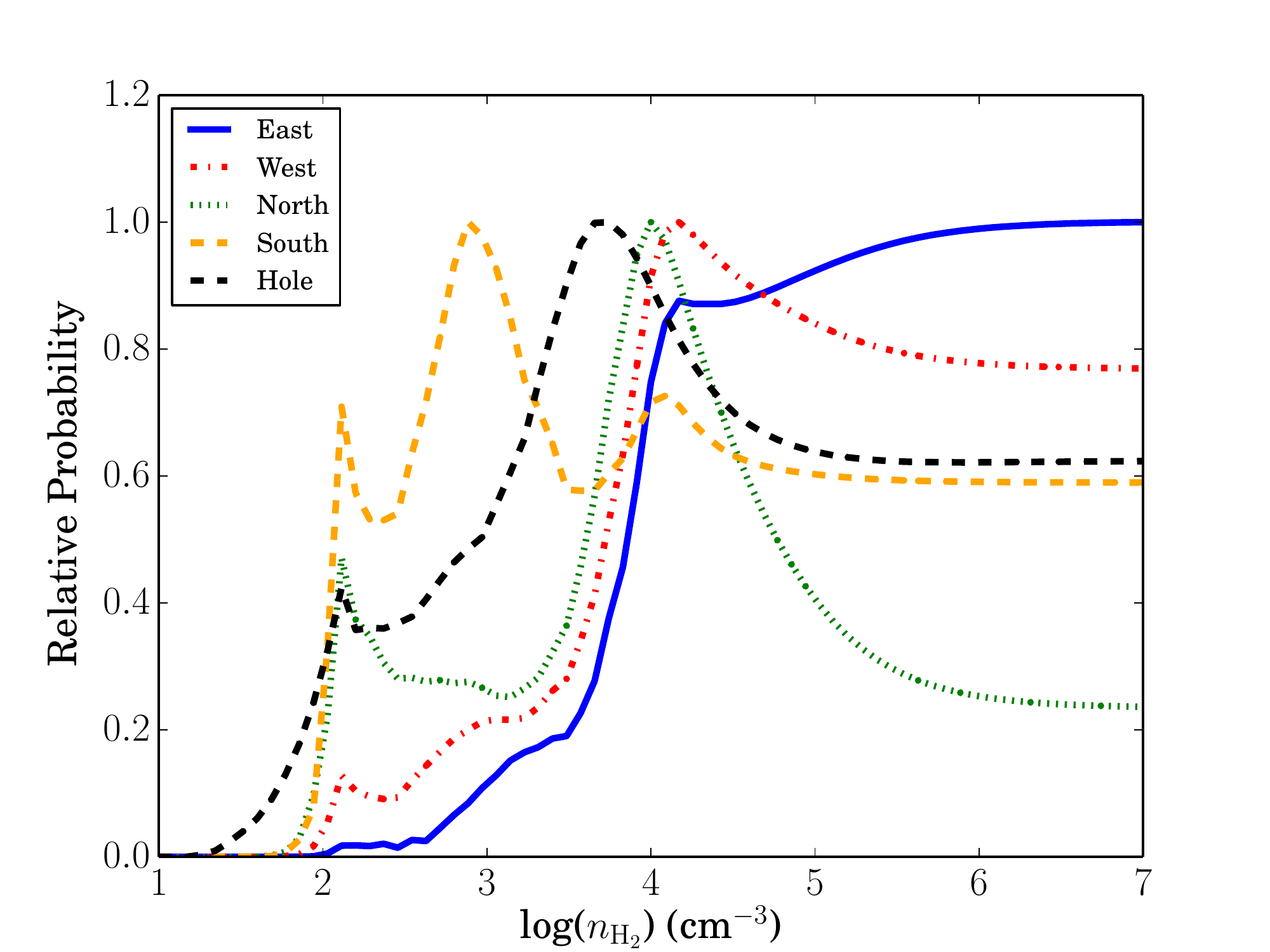} \\
\includegraphics[scale=0.4]{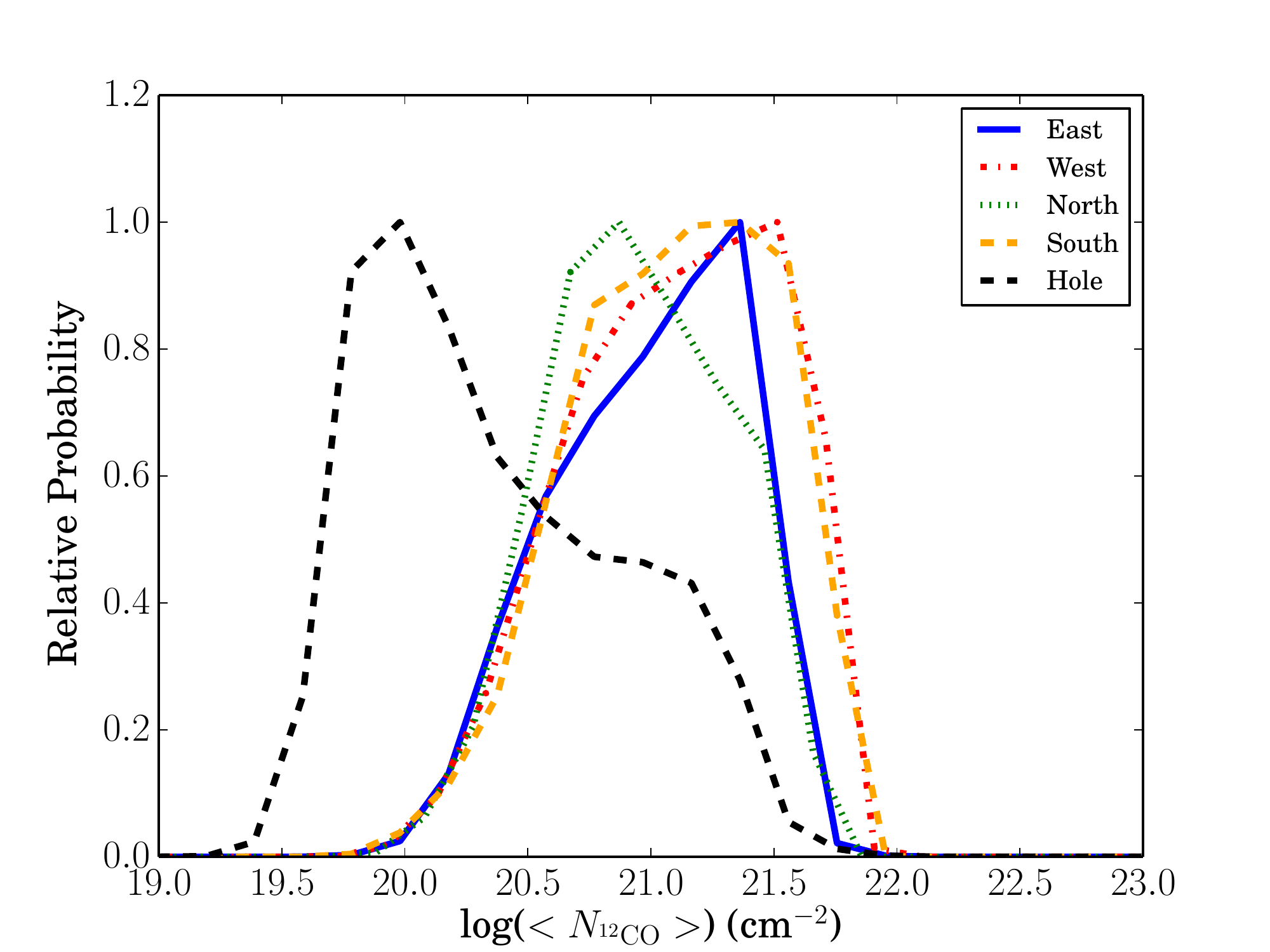}  & \includegraphics[scale=0.4]{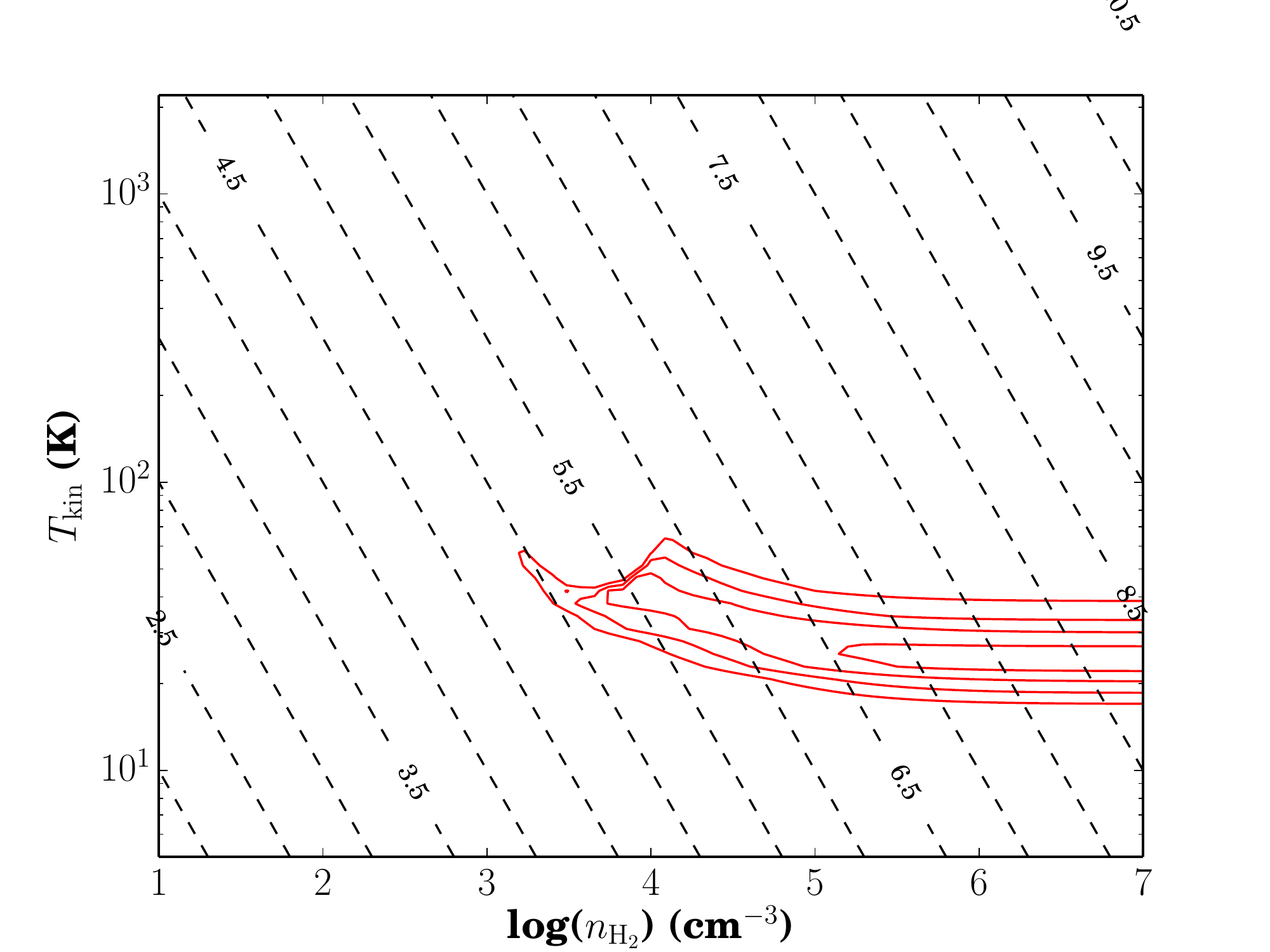} \\
\includegraphics[scale=0.4]{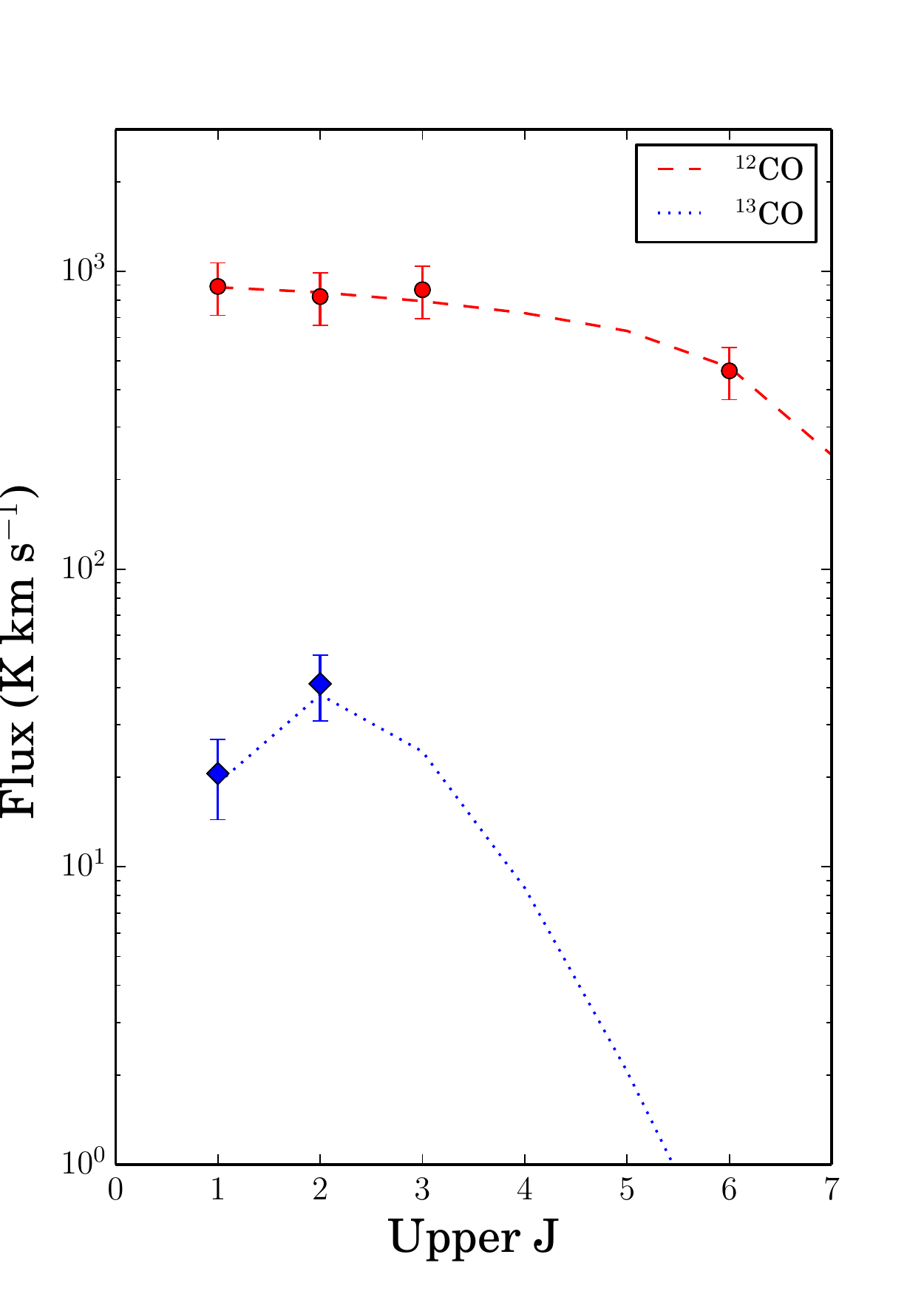}  & \includegraphics[scale=0.4]{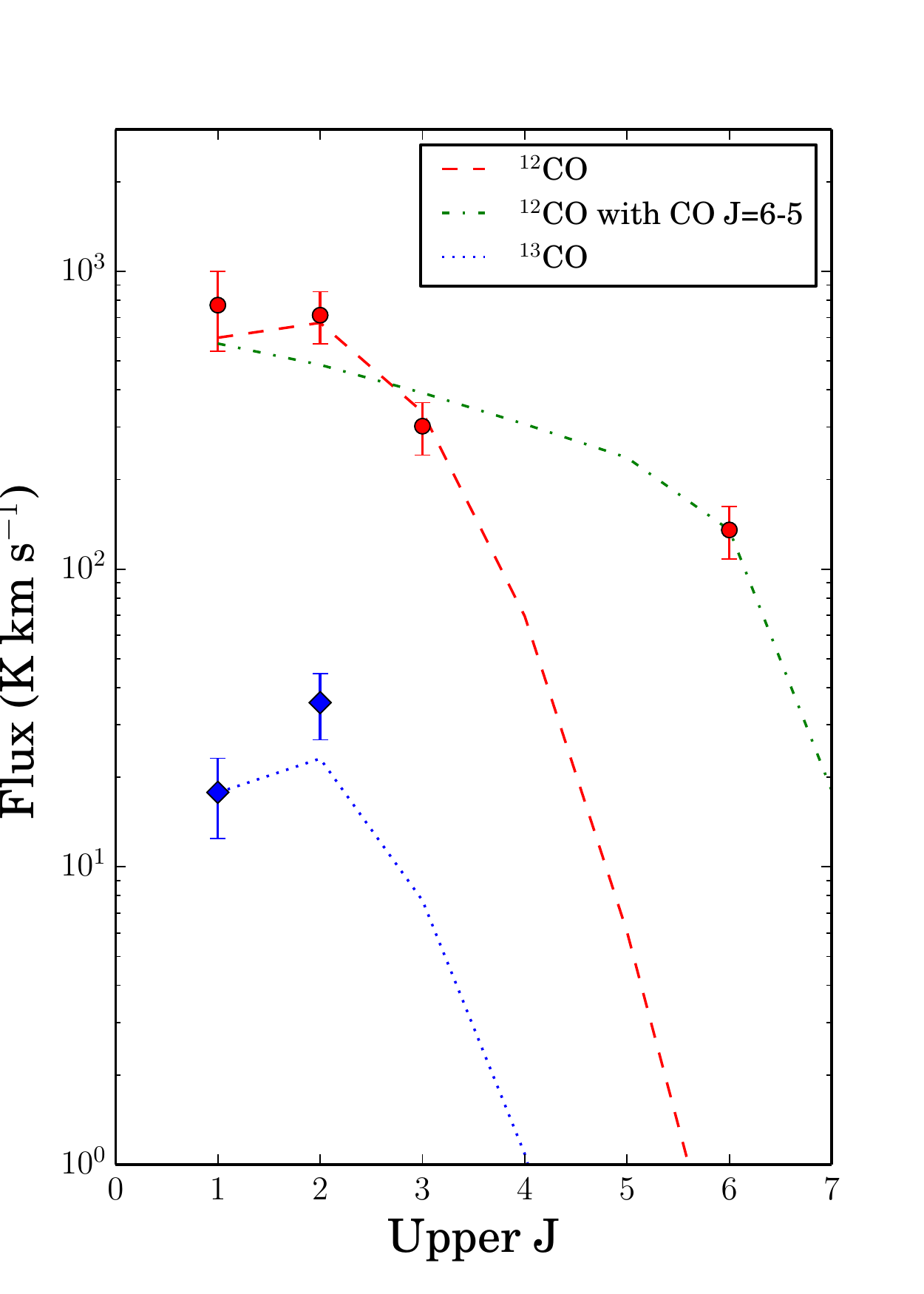} \\

\end{array}$

\caption[]{
\textit{Top Row:}  (Left) Probability distribution of \tkin. (Right) Probability distribution of log(\nhtwo).
\textit{Middle Row:}  (Left) Probability distribution of log($<$\nco$>$). (Right) Sample probability contour plot of \tkin\ versus log(\nhtwo) for the eastern region. Solid contours represent (30,50,70,90)$\%$ of the maximum probable solution. Dashed contours represented pressure.
\textit{Bottom Row:}  (Left) Example spectral line energy distribution (SLED) for the eastern region of NGC 1614. Dashed lines represent the 4DMax solution. (Right) CO SLED for the ``hole". Green dashed lines represent the 4DMax solution that includes the \cosix\ flux into the modelling. 

 }
\label{probdist}
\end{figure}

\newpage
\begin{deluxetable}{cccccc} 
\tablecolumns{5}
\tablewidth{0pt}
\tablecaption{Data for NGC 1614}
\label{smasum}
\tablehead{\colhead{Transition Line} & \colhead{Observatory} & \colhead{Interferometric Flux\tablenotemark{a}}  &\colhead{Resolution} & \colhead{rms \tablenotemark{b}} \\ 
\colhead{} & \colhead{} &  \colhead{(Jy \kms)}   & \colhead{(\arcsec)} & \colhead{(mJy beam$^{-}$)} }
\startdata
\coone\	& CARMA & 240 $\pm$ 14  	 	&3.3 x 2.9 	& 4 \\
\cotwo\ 	& SMA 	& 1250  $\pm$ 5  		&0.7 x 0.6 	& 7.5\\
\cothree\ & SMA 	& 1160 $\pm$ 80  	 	&2.7 x 2.2 	& 43\\
\cothree\ & ALMA 	& 1680 $\pm$ 10  	 	&0.7 x 0.4 	& 3 \\
\cosix\ 	& ALMA 	& 920 $\pm$ 10  	 	&0.4 x 0.3 	& 7 \\
\hcofour\ 	& ALMA 	& 24.3 $\pm$ 0.6 &  0.6 $\times$ 0.4 & 0.9 \\
\hline 
\tcoone\ 	& CARMA & 3.6 $\pm$ 0.5  	 	&3.8 x 3.2 	& 3.1 \\
\tcotwo\ 	& SMA	 & 16.0 $\pm$ 0.5  	 	&0.8 x 0.7  	& 3.3 \\
\hline
Continuum (435$\mu$m$_{rest}$) &ALMA&  240 $\pm$ 4 (mJy)  & 0.3 x 0.2 & 1.4\tablenotemark{c} \\

\enddata
\tablenotetext{a}{Measurement uncertainty only. Calibration uncertainty is 20$\%$ for all maps.}
\tablenotetext{b}{Velocity channel widths of 20 \kms\ for \co\ and HCO$^{+}$ maps and 100 \kms\ for \tco\ maps.}
\tablenotetext{b}{Noise level determined using the combined USB and LSB map.}

\end{deluxetable}

\begin{deluxetable}{@{\hspace{-0.3in}}c@{\hspace{-0.3in}}cccccccc} 
\centering
\tabletypesize{\scriptsize}
\tablecolumns{9}
\tablewidth{0pt}
\tablecaption{Modelling Results}
\label{modelresults}
\tablehead{\colhead{Region} & \colhead{} & \colhead{\tkin\ } & \colhead{\nhtwo\ }  & \colhead{$Pressure$} & \colhead{$<$\nco\ $>$} & \colhead{Mass} & \colhead{\xco\ } & \colhead{$\alpha_{CO}$ }  \\
\colhead{} & \colhead{} & \colhead{(K)} & \colhead{(cm$^{-3}$) }  &   \colhead{(K cm$^{-3}$)} & \colhead{(cm$^{-2}$)} & \colhead{(\msol)} & \colhead{ }  & \colhead{ (\alphaco)}   }
\startdata
North& 1DMax & 25 & 10$^{4.00}$ &   10$^{5.47}$ & 10$^{18.87}$ & 10$^{7.33}$ & 131 &  1.2\\
 	& 4DMax &  31 & 10$^{3.83}$ & 10$^{5.47}$ & 10$^{18.87}$ & 10$^{7.33}$ & 131 &	    1.2	\\
 	& 1$\sigma$ range & 21-64 & 10$^{2.98}$-10$^{5.67}$ & 10$^{5.05}$-10$^{7.11}$ & 10$^{18.44}$-10$^{19.27}$ & 10$^{6.90}$-10$^{7.73}$ & 83-575 & - \\
\hline
West& 1DMax & 23 & 10$^{4.17}$ &   10$^{5.86}$ & 10$^{19.51}$ & 10$^{7.98}$ & 575 & 1.2\\
 	& 4DMax &  34 & 10$^{3.74}$ & 10$^{5.86}$ & 10$^{19.51}$ & 10$^{7.98}$ & 575 & 1.2\\
 	& 1$\sigma$ range & 20-41 & 10$^{3.90}$-10$^{6.37}$ & 10$^{5.43}$-10$^{7.85}$ & 10$^{18.53}$-10$^{19.47}$ & 10$^{6.99}$-10$^{7.93}$ & 104-954 & -  \\
\hline
South& 1DMax & 23 & 10$^{4.17}$ &  10$^{5.86}$ & 10$^{19.36}$ & 10$^{7.82}$ & 575  &0.9\\
 	& 4DMax &  34 & 10$^{3.83}$ & 10$^{5.86}$ & 10$^{19.36}$ & 10$^{7.82}$ & 575 & 0.9\\
 	& 1$\sigma$ range & 19-39 & 10$^{3.90}$-10$^{6.39}$ & 10$^{5.43}$-10$^{7.83}$ & 10$^{18.47}$-10$^{19.44}$ & 10$^{6.93}$-10$^{7.90}$ & 102-1020 &- \\
\hline
East& 1DMax & 25 & 10$^{7.00}$ &   10$^{8.28}$ & 10$^{19.36}$ & 10$^{7.82}$ & 691 & 1.5\\
 	& 4DMax &  42 & 10$^{3.74}$ & 10$^{8.28}$ & 10$^{19.36}$ & 10$^{7.82}$ & 691  & 1.5\\
 	& 1$\sigma$ range & 20-41 & 10$^{4.18}$-10$^{6.49}$ & 10$^{5.76}$-10$^{8.02}$ & 10$^{18.45}$-10$^{19.32}$ & 10$^{6.92}$-10$^{7.78}$ & 148-1122  &  -\\
\hline
Hole& 1DMax & 11 & 10$^{3.74}$ &   10$^{4.96}$ & 10$^{17.98}$ & 10$^{6.44}$ & 52 & 0.1\\
 	& 4DMax &  15 & 10$^{3.91}$ & 10$^{4.96}$ & 10$^{17.98}$ & 10$^{6.44}$ & 52 & 0.1 \\
 	& 1$\sigma$ range & 10-59 & 10$^{3.04}$-10$^{6.16}$ & 10$^{4.77}$-10$^{7.43}$ & 10$^{17.72}$-10$^{18.90}$ & 10$^{6.19}$-10$^{7.36}$ & 42-489 & -\\\hline

\enddata

\end{deluxetable}

\end{document}